\documentclass[twocolumn]{aa}

\date{Received date / Accepted date }

\usepackage[varg]{txfonts}
\usepackage[utf8]{inputenc}
\usepackage[english]{babel}
\usepackage{tabularx}

\usepackage{setspace}

\usepackage{indentfirst}

\usepackage{enumitem}
\setlist[itemize]{noitemsep, topsep=0pt}

\usepackage{graphicx}

\graphicspath{{/Users/elka127/Documents/uio/aa-paper}}

\usepackage{wrapfig}
\usepackage{subfigure}

\usepackage[dvipsnames]{xcolor}

\usepackage{hyperref}
\hypersetup{
    colorlinks,
    urlcolor=blue,
    linkcolor=blue,
    citecolor=blue
}

\usepackage{csquotes}

\usepackage{listings}

\usepackage{courier}

\definecolor{bluekeywords}{rgb}{0,0,1}
\definecolor{greencomments}{rgb}{0,0.5,0}
\definecolor{redstrings}{rgb}{0.64,0.08,0.08}
\definecolor{types}{rgb}{0.17,0.57,0.68}
\lstdefinestyle{py}
{
    language=Python,
    frame=l,
    framesep=5pt,
    captionpos=b,
    numbers=left,
    numberstyle=\tiny,
    showspaces=false,
    showtabs=false,
    breaklines=true,
    showstringspaces=false,
    breakatwhitespace=true,
    commentstyle=\color{greencomments},
    keywordstyle=\color{bluekeywords},
    stringstyle=\color{redstrings},
    basicstyle=\footnotesize\ttfamily,
}
\newcommand\Tstrut{\rule{0pt}{2.6ex}}       
\newcommand\Bstrut{\rule[-0.9ex]{0pt}{0pt}} 
\newcommand{\TBstrut}{\Tstrut\Bstrut} 
\usepackage{natbib}
\bibpunct{(}{)}{;}{a}{}{,} 
\begin{document}

\title{Planets similar in size are often dissimilar in interior}

\author{E. Mamonova\inst{1}\fnmsep\inst{2}, Y. Shan\inst{1}\fnmsep\inst{2}, P. Hatalova\inst{1}\fnmsep\inst{2}, S.C. Werner\inst{1}\fnmsep\inst{2}}

\institute{Centre for Planetary Habitability (PHAB), University of Oslo, 0315 Oslo, Norway
\and Centre for Earth Evolution and Dynamics (CEED), University of Oslo, 0315 Oslo, Norway}
\abstract{The number of discovered exoplanets now exceeds 5500, allowing statistical analyses of planetary systems. Multi-planet systems are mini-laboratories of planet formation and evolution, and analysing their system architectures can help us to constrain the physics of these processes. Recent works have found evidence of significant intrasystem uniformity in planet properties such as radius, mass, and orbital spacing, collectively termed `peas in a pod' trends. In particular, correlations in radius and mass have been interpreted as implying uniformity in planet bulk density and composition within a system. However, the samples used to assess trends in mass tend to be small and biased. In this paper, we re-evaluate correlations in planet properties in a large sample of systems with at least two planets for which mass and radius have been directly measured, and therefore bulk density can be calculated. Our sample was assembled using the most up-to-date exoplanet catalogue data, and we compute the relevant statistics while using a procedure to `weight' the data points according to measurement precision. We find a moderate correlation in radius and a weak correlation in the densities of adjacent planets. However, masses of neighbouring planets show no overall correlation in our main sample and a weak correlation among pairs of planets similar in size or pairs restricted to M$_p$<100 M$_\oplus$, R$_p$<10 R$_\oplus$. Similarly, we show that the intrasystem dispersion in radius is typically less than that in mass and density. We identify ranges in stellar host properties that correlate with stronger uniformity in pairs of adjacent planets: low T$_\mathrm{eff}$ for planet masses, and low metallicity and old age for planet densities. Furthermore, we explore whether peas in a pod trends extend into planet compositions or interior structures. For small neighbouring planets with similar radii, we show that their masses and interior structures are often disparate, indicating that even within the same system, similarity in radii is not necessarily a good proxy for similarity in composition or the physical nature of the planets.
}
\titlerunning{Planets similar in size are often dissimilar in interior}
\authorrunning{Mamonova et al.}
\maketitle

\section{Introduction}
\label{sec:introduction}
Since the 1995 discovery of the first exoplanet orbiting a main sequence star \citep{QuelozDidier1995AJct}, there has been explosive growth in the number of detections of these objects. The results from the Kepler mission provided the first statistical picture of the properties of planetary systems \citep{borucki2010kepler}, and we can compare the Solar System with other star systems in our galaxy. The transit method used in this mission, and its extension, the NASA K2 mission \citep{howell2014k2}, contributed to the determination of fundamental planet parameters such as radius and orbital period for more than 3500 planets. Follow-up radial velocity (RV) surveys and observed transit timing variations (TTVs) have helped to determine planet masses in an ever-growing number of systems.

In the early days of exoplanet exploration, the main challenge was to detect the gravitational acceleration of a small planet induced on its host stars or other planets in the system with the available ground or orbital instruments. Therefore, the idea of the theoretical mass--radius relation (MRR) was appealing both for planet mass prediction and the development of the first theories of observed exoplanet interiors and atmospheres. The MRRs are typically based on exoplanet data collected in the main exoplanet catalogues, such as observed mass and radius distributions, and are designed to explain these observations. Recent studies \citep{chen2016probabilistic, weiss2013mass, weiss2014mass, Wolfgang2016, bashi2017two, otegi2020revisited} used different approaches, such as power-law fitting to the whole known population, or suggesting transition regions in the populations.

Initial analyses of available data indicated a rich diversity in observed exoplanets \citep{santerne2019extremely, armstrong2020remnant, hoeijmakers2018atomic,winn2018kepler, espinoza2020hd, demory2016map, evans2016detection}. It was soon established that the MRR diagram presents a large scatter, making it difficult to disentangle subsets in the population. The power-law fitting and other simplified methods of producing MRRs began to face increasing difficulties in explaining the observed diversity. A different approach to suggesting a specific MRR is to take into account thermal equilibrium calculations and to predict the likely occurrence of particular species and mineralogy within the interior of a planet, along with their equations of state (EOS). This approach was implemented by \citet{zeng2016mass,zeng2019growth}, who calculated the MRR curves for prescribed planet bulk compositions, including compression. Both the Kepler and ongoing TESS \citep{ricker2016transiting} mission discovered a large population of super-Earths and mini-Neptunes, sometimes at close-in orbital distances. The lack of a planet of analogous size (and possible composition) in the Solar System means that we cannot fully rely on knowledge of internal planetary structures in our neighbourhood to interpret the majority of known exoplanets.

While radius is the most straightforward to measure, the density and interior structures of exoplanets hold key information about their physical nature and origins. \citet{2012Sci...337..556C}, who present an example of a Kepler-36 system with the smallest fractional separation between orbits of two planets and the largest density contrasts between them, raised interesting questions. If the two planets formed in situ, the compositions and densities of the planets could have changed with time, for example because of the erosion of the  atmosphere of the closest planet by instellation. On the other hand, formation at two widely separated locations in the protoplanetary disc, for example in the volatile-poor and volatile-rich regions, would suggest a `migration' mechanism capable of drawing two planets together. \citet{2022Sci...377.1211L} proposed a density gap that separates rocky from water-rich exoplanets and their results are inconsistent with the previously known bimodal radius distribution arising from atmospheric loss of a hydrogen and helium envelope. These considerations call for a focus on planet density and interior studies.

To understand the interiors and constituents of exoplanets, it would be helpful not only to base our theories on mass and radius determinations of  planets but also to gather insights from a planetary system as a whole. Exoplanet observations indicate a wide variety in the architecture of multi-planet systems \citep{lissauer2011architecture, fabrycky2014architecture, winn2015occurrence}, but most known systems have planets at separations of less than the orbit of Mercury that are simultaneously much larger and heavier than our innermost planet. The Solar System was once again found to be quite different from the others. Furthermore, the observations hint at the existence of several intrasystem correlations \citep{millholland2021split, otegi2021similarity} also not observed in our system. Exoplanets within an exoplanetary system tend to be of similar size and mass and evenly spaced, and are often ordered in size and mass, as was first suggested by \citet{weiss2018california}, \citet{Wang_2017}, and \citet{millholland2017kepler}. Similarity implies that $P_{i+1}/P_i$ $\approx$ 1, where $P$ is a chosen planet parameter, which can be mass, radius in pairs of adjacent planets, or period ratios for the second and first pair in triples. However, the authors of the several papers mentioned above, and of some mentioned below, formulated these trends differently using different restrictions posed on their samples, and sometimes looked only for similarity in one or two parameters. \citet{weiss2018california} found the correlation between the size of a planet compared to the next planet in the system in order of increasing orbital period for the sample, which they constructed solely from the Kepler photometry using both confirmed planets and planet candidates, but posing multiple restrictions on the accuracy of the determined planet radius in order to work with a high-purity planet sample. These latter authors also found similarity in the period ratio in the sample restricted to period ratios of smaller than 4. \citet{Wang_2017}, working with a sample of 29 systems with at least three RV-detected planets, found 15 systems with planets with M$_p$<30 M$_\oplus$ that show unusual similarity in mass. Most of their sample consists of non-transiting planets, which prevented similarity analyses in radius. Analysing a small sample of 89 planets with masses determined by TTVs only, \citet{millholland2017kepler} found that planets in a system have both masses and radii that are significantly more similar than in the mock systems assembled from randomly shuffling planets in their sample. \citet{otegi2021similarity} reported significant similarity in radius, mass, and density in planet pairs belonging to the same system, especially if the system hosts planets with M$_p$<100 M$_\oplus$.  These authors found that planetary systems tend to be more similar in radius than in mass, and that generally, the planets most similar in radius do not correspond to those similar in mass.

Together, all spotted correlations were collectively referred to as `peas in a pod' trends in the architecture of planetary systems. These trends jointly suggest that planets similar in size can have similar bulk densities (and perhaps similar chemical compositions). The physical reality of these trends has been challenged and attributed to observational biases, for instance, arising from Kepler's discovery efficiency at the detection threshold (\citet{zhu2020patterns}, but see \citet{weiss2020kepler}).
\citet{murchikova2020peas} showed that apparent patterns can arise in radius distribution models with simple assumptions, without requiring planet radii to be correlated between neighbours.
On the other hand, some of these correlations have been reproduced in planet population simulations \citep{2019MNRAS.490.4575H,2020AJ....160..276H,mishra2021new}. \citet{2019MNRAS.490.4575H} found that models that assume planets have independent periods and sizes are inconsistent with the observationally derived properties of multi-planet systems in the Kepler sample. These latter authors suggest a clustered model where planets in the system tend to be similar in radius within the cluster and periods are drawn conditionally on the period scale of the parent cluster. They restricted their analyses to modelling planets with R$_p$<10 R$_\oplus$, and in order to adequately simulate mock systems, they drew planet masses using the MRR from the non-parametric model in \citet{2018ApJ...869....5N}. \citet{2019MNRAS.490.4575H} did not analyse the masses for similarity. The same authors later showed that the observed patterns in the ordering of planet sizes and uniform spacings cannot be produced by the detection biases of the Kepler mission  \citep{2020AJ....160..276H}. Using a modern planet population-synthesis model and forward modelling of the detection biases and completeness of the Kepler survey, \citet{mishra2021new} revealed the peas in a pod trends present in their simulations. These authors produced samples of planets with masses of up to 10$^4$ M$_\oplus$ and showed that both their models ---one that is unrestricted by observational biases and one that simulates the Kepler observational field--- exhibit a high level of correlation in adjacent planet radii (R $\approx$ 0.64-0.75) and in mass (R $\approx$ 0.6-0.76). The models also exhibit a weak (R= 0.25) to moderate (R=0.55) correlation in period ratios for restricted and unrestricted simulated populations, respectively. They formulated these trends as such:

– Similarity in size: planets within a system tend to be either similar in size, in the sense described above, or ordered in size.

– Similarity in mass: planets within a system tend to be either similar or ordered in mass.

– Similarity in spacing: for systems with three or more planets, the spacing between a pair of adjacent planets is similar to the spacing between the next pair of adjacent planets.

The main challenge for evaluating the peas in a pod trend as an astrophysical feature could arise from the incompleteness of the observed exoplanet population. First of all, at present, and with sufficient precision, we can determine the radii of transiting planets only \citep{1952Obs....72..199S}. Further, the Kepler mission, which was for many years the most productive mission in providing fundamental exoplanet parameter data, had its observational limitations, especially for small planets, due to transit depth measurement constraints and difficulties in detecting exoplanets at orbital periods of greater than 200 days. As planet occurrence increases towards small size, and Earth-sized planets are very common (e.g. \citet{petigura2013prevalence,fulton2017california,burke2015terrestrial,mulders2018exoplanet}), it is likely that at least some of the known systems harbour additional undetected planets that could skew the observed peas in a pod trends.

The other important factor that could affect trends emerging from the known population is the accuracy of measurements of exoplanet parameters such as mass and radius. The determination of these  parameters relies on measurements of the host star radius and mass and can be imprecise and inhomogeneous. Recently released results of the Gaia astrometric space mission (Gaia DR3), which provide the astrometry, photometry, and dispersed light for hundreds of millions of sources from the Gaia prism spectra and the spectrograph \citep{fouesneau2022gaia}, provide a homogeneous set of stellar radii. Initially, in this study, we worked with planet parameters retrieved from the exoplanet databases as described below in Section \ref{sec:methods}. We subsequently re-evaluated the radius trends using homogeneously derived stellar radii from the Gaia DR3.

Intrasystem correlations in planet radii and mass can reflect the degree of uniformity in the internal structures and compositions of planets within the same system. One might expect planets born in the same system to have a similar composition, especially if they are formed near one another. After all, the terrestrial planets in our Solar System and the TRAPPIST-1 system are remarkably uniform \citep{agol2021refining}. Indeed, \citet{otegi2021similarity} found a strong correlation in bulk density between adjacent planets in multi-planet systems, especially around M dwarfs. However, bulk density is a simplistic proxy for interior composition and does not fully capture mass-dependent self-compression and pressure--temperature-related phase transitions and material changes. In a recent study, \citet{2023AJ....166..137R} investigated the core and water mass fractions for multi-planet systems orbiting M dwarfs and found no evidence of intrasystem uniformity (and found a hint of significant non-uniformity in some cases). Though the sample studied by these latter authors is small, their work indicates that at least some stars host rocky and volatile-rich planets side by side. It would be interesting to extend this analysis to a larger sample of carefully vetted multi-planet systems.

In the present paper, we revisit the peas in a pod trends, as formulated by \citet{mishra2021new}, in the exoplanetary system architecture, using recent exoplanet and stellar data. We aim to assess whether or not the similarity of a pair of planets in one parameter (e.g. radius) can be a good predictor of their similarity in other parameters (e.g. mass or density). The paper is organised as follows: in Section \ref{sec:methods}, we explain the process of assembling this study sample and our methods of analysing the uniformity of planetary systems in planet parameters. In Section \ref{sec:results}, we then report our findings in terms of the patterns we see in mass, radius, density, and period spacing. Our conclusions are discussed in Section \ref{sec:discussion}. and are summarised in Section \ref{sec:conclusions}.

\section{Methodology}
\label{sec:methods}

\subsection{The sample}

The data for this study were initially queried from \href{https://exoplanetarchive.ipac.caltech.edu}{The NASA Exoplanet Archive} \citep{akeson2013nasa}, which was accessed on July 18, 2022. We cross-checked the retrieved parameter values with the \href{https://exoplanet.eu}{Extrasolar Planets Encyclopaedia} \citep{schneider2011defining}. Additionally, the methods of mass detection for planets in the sample were retrieved from papers that reported this parameter. The reported masses were cross-validated between the two databases in order to exclude parameters obtained from theoretical MRR. The majority of this study's sample came from the Kepler and TESS missions. In order to explore the architecture of the known planetary systems, and to study them as a family of objects, it is critical to assemble a sample consisting of multi-planet systems, where the configuration of the system is at least partially established.

We assembled an initial sample of 528 exoplanets in 175 multi-planet systems by selecting exoplanets from the systems where at least two planets have measurements of mass and radius. Planets discovered by the direct imaging method, which typically probes young A-type stars and very large planets on wide orbits, often have conflicting reported parameters, such as mass and radius. Supermassive planets usually come from young unstable, often multi-star systems, and their parameters are observed with large uncertainties \citep{galicher2016international}. HR 8799, PDS 70, and TYC 8998-760-1 systems formally made it to the sample but were eliminated for these reasons. Therefore, the mass and density samples include only planets with masses measured by RV or TTVs, and the radius sample includes planets with radii measured from transits. In order to analyse planetary systems architecture, we include in our sample all known parameters of the host star and all planets. We then extracted separate samples for adjacent inner and outer planet parameters, if two consecutively located planets have them reported. For the analyses of features in the systems architecture such as mass and size distribution, and spacing, the subsamples were further vetted to include only the systems containing at least two adjacent planets with the known parameters of choice ($P$) and which have the sum of uncertainties $\sigma_{P\mathrm{sum}}\leq$ 2$P$ (i.e. $\sigma_{P\mathrm{sum}}$ is equivalent of 2$\sigma$, where $\sigma$ is standard deviation). For period ratio analyses, systems from our main sample with at least three planets were included. For studying an intrasystem parameter's similarity, we include all planet pairs or triples belonging to the same system.

We started with forming separate subsamples for masses (184 pairs), radii (269 pairs), calculated bulk densities (167 pairs) and period ratios (167 triples) of adjacent planets by filtering the data to consider only exoplanets satisfying our uncertainty criteria. Additionally, planets with uncertainty flags\footnote{Uncertainty flag appears if only lower or upper limit of a value was reported.} were excluded. In our subsamples, if the planetary system has more than 2 planets, all formed planet pairs are included in the correlation calculation, representing multiple data points. To further address using uncertainties in parameters in our sample, slightly larger than usual, we compared our methods with ones used by \citet{otegi2021similarity}, and only a handful of planets in our sample are above the threshold suggested by them (see Fig.\ref{fig:33}, left in Appendix \ref{sec:appendix}). As we demonstrate below in Section \ref{sec:results}, our outcomes were largely not affected even when the sample was further restricted for uncertainties even smaller than that.

It should be noted that as with any other, this study catalogue suffers from observational biases, and database errors and is essentially incomplete. In Section \ref{sec:results}, we explore the effect of inhomogeneity in stellar parameter measurements on the results by constructing a new sample of planet radii. General Stellar Parametrizer from Photometry (GSP-Phot) \citep{recio2022gaia}, which is designed to produce a homogeneous catalogue of parameters for hundreds of millions of single non-variable stars based on their astrometry, photometry, and low-resolution BP/RP spectra, provides stellar radii \citep{andrae2022gaia}. Alternatively, Final Luminosity Age Mass Estimator (FLAME) \citep{2022arXiv220605864C} estimate the stellar radii based on the stellar astrophysical parameters from either GSP-Phot or General Stellar Parametrizer from Spectroscopy (GSP-Spec) combined with the Gaia photometry and parallaxes. We retrieved the radii of the sample stars from FLAME and GSPPhot from Gaia DR3, which we retrieved from the \href{https://gea.esac.esa.int/archive/}{Gaia Archive}, and calculated new radii using planet-to-star ratios that were queried from The Q1-Q17 Data Release 25 (DR 25) Kepler Objects-of-Interest (KOI) \citep{hoffman2017uniform} if available, otherwise, the most recent values were queried from the NASA Exoplanet Archive. We used this new radii sample to check if the results of our analyses would differ from those performed on our initial sample. To address inhomogeneity in mass in our sample, we identify the method of mass determination for all planets in the sample with known mass, using both The NASA Exoplanet Archive and Extrasolar Planets Encyclopaedia. We checked that the distributions of planets with TTVs and RV-determined masses are similar in mass-radius space, see  Fig.\ref{fig:33} in Appendix \ref{sec:appendix}, right. In this Figure, all mass bins from smaller than 1M$_\oplus$ up to  10$^3$M$_\oplus$ have planets whose masses were detected either with one or another method. There are no visible clusters in the distribution of planets’ mass determined by these two methods, and therefore, we assume any method-related mass bias to be small.

As a result, we reported our findings based on the best and most recent data to our knowledge. The data of the planets included in our main sample catalogue is accessible in the supplementary materials in ASCII format\footnote{The table with the main sample planet parameters is only available in electronic form at the CDS via anonymous ftp to \url{cdsarc.u-strasbg.fr} (130.79.128.5) or via \url{http://cdsweb.u-strasbg.fr/cgi-bin/qcat?J/A+A/.}}. The code prepared here for assembling and analysing the sample can be found on GitHub\footnote{\url{https://github.com/cepylka/exoplanets-system-architecture}}.

\subsection{Statistical analyses}

The Pearson R test measuring the strength of correlations between two variables \citep{pearson1905general} is a powerful method to investigate the correlation in two distributions.  R values vary between -1 and +1 with zero implying no correlation and $\pm$1 implying an exact linear relationship. The probability that an uncorrelated sample of the same size will show the same R is denoted by the p-value and indicates the significance of the results. The test assumes a normal distribution and a linear relationship between two variables, and therefore suffers a setback for small samples (N<30) \citep{jiang2020orbital}. Non-normality caused the correlation coefficient inflation, particularly in the case of extremely skewed distributions, but for large samples, this effect is less prominent \citep{bishara2015reducing}. The Pearson test can also fail if the dataset consists of data points gathered around the unity line at the low and high tail ends. If the pairs of very similar values belonging to the two populations, comparing in the test, occupy the high and low tails of distributions, it would lead to apparent correlation, and in the opposite, highly dissimilar values on the one tail of distribution would `smear' out the possible correlation. Additionally, the Pearson correlation coefficient as a method has its limitations for searching for correlations in sufficiently large samples (typically N$\gtrsim$ 500) or insufficiently small samples (N$\lesssim$ 30), such that the significance levels of Pearson’s R may be biased due to the sensitivity of p-value to N. Inherently, the p-value calculated along with Pearson R as a function solely depends on sample size and value of R, and can therefore be very small with large R and N. The confidence intervals, however, can be a reality check, showing the interval where the actual value of R might be at a certain degree of confidence. That is why we calculated them for R using Fisher's $z'$ transformation \citep{10.2307/2331838}, which is used due to the non-normality of the distributions.

Nevertheless, this method was suggested by multiple studies for finding a correlation in fundamental parameters of two adjacent planets within observed systems \citep{weiss2018california, jiang2020orbital, millholland2021split, otegi2021similarity}. Therefore, we also used the Pearson correlation test in this study and further discussed its caveats in Section \ref{sec:discussion}.

First, we performed this test in the initial sample, comparing planet parameters such as mass, radius, density in planet pairs and period ratio in triples. In forming our samples, we chose to maximise the number of planets included, therefore imposing an uncertainty threshold that is larger than previous authors for similar studies (for example, \citet{otegi2021similarity} suggested as a cut-off $\sigma_{m\mathrm{sum}}\leq$ 1$m$ and $\sigma_{r\mathrm{sum}}\leq$ 0.32$r$ for mass and radius uncertainties, respectively). The number of planets in our sample with uncertainties larger than in previous studies is relatively small (see Fig.\ref{fig:33} in Appendix \ref{sec:appendix}). To account for the impact of large uncertainties, we implemented a routine that `weights' the data points according to the measurement precision by performing multiple (10$^5$) Monte Carlo experiments when the results are computed based on a repeated random uniform sampling of data points ($x,y$) from the observational error intervals ($x \pm dx, y\pm dy$). At every step during the simulations, we also performed the Pearson correlation test.
Then we performed subsequent statistical analysis to determine the median values of the distribution. These median values can be considered the true values of the correlation coefficients. We note that the computation of the Pearson coefficients initially does not include the uncertainties, but our weighting or `error-accommodation' technique,  allows us to overcome this without complicated calculations.

To establish the null hypothesis as a comparison, we constructed test populations from the non-parametric model in \citet{2018ApJ...869....5N} which relates the planet’s radius to its mass. Using observed radii in our sample, we randomly draw test populations of mass and density in such a way, that any adjacent pair of planets do not violate stability in the system. For that, we assumed that mock planets occupy the same orbits as in the real systems, and calculated stability criteria as formulated in \citet{deck2013first}: in the case of initially circular orbits two bodies in a planetary system should satisfy the following equation for semi-major axes:

\begin{equation}
    \frac{a_2 - a_1}{a_1} \gtrsim 2.4 \epsilon_p^{1/3},
    \label{eq:2.7}
\end{equation}
where $\epsilon_p = (M_{p(i)}+M_{p(i+1)})/M_\star$ is the total mass of the planets relative to the mass of the host star. This equation was used to determine if we produced any unstable planet pairs, and if so, we eliminated them from the calculation. This way of constructing the mock populations allowed us to preserve observed structure in the most precisely up-to-date determined fundamental parameters of exoplanet: in radius and period. After acquiring 10$^5$ test populations, we performed statistical analyses and calculated mean or median values (in case of heavily skewed distributions). We used the test population to compare with the Pearson test results and consequently in further analyses of the sample.

Another way to address the intrasystem similarity as a trend is an evaluation of the mean normalised parameter dispersion (first formulated by \citet{2022AJ....163..201G} as adopted a modified version of the mass uniformity metrics from \citet{Wang_2017}, and subsequently used by \citet{2023MNRAS.525L..66L}):

\begin{equation}
   D = \frac{1}{N_{sys}} \sum_{i=1}^{N_{sys}}\frac{\sigma_{p_i}}{\overline{p_i}} = \frac{1}{N_{sys}}  \sum_{i=1}^{N_{sys}}\sqrt{\frac{\sum_{j=1}^{N_{pl_i}}(p_j - \overline{p_i})^2}{\overline{p_i}^2(N_{pl_i} - 1)}},
  \label{eq:2.1}
\end{equation}
where $N_{pl_i}$, $\overline{p_i}$, and $\sigma_{pi}$ are the number of planets, mean parameter value, and parameter's standard deviation in the $i$-th system, respectively. If the value of $D$ is zero, the parameter's dispersion is zero, and with rising values of $D$ similarity in the system dissipates. Essentially, $D$ represents the sum of standard deviations divided by means or coefficients of variation of parameters in all systems averaged by a number of systems. For example, a $D$= 0.5 signifies that the system-wide spread in the parameter of interest is on average half the mean value of that parameter in the given system. We calculate intrasystem parameter dispersion $\frac{\sigma_{p_i}}{\overline{p_i}}$ for masses, radii, densities and period ratios for each system in our sample, and $D$ for these parameters for the entire sample and several subsamples.

The fundamental parameters of the protostar can lead to formation of planets with similar properties. For example, \citet{millholland2021split} suggested that low-metallicity stars are harbouring planets more equal in size. We probed the hypothesis, that subsets of data points are correlated stronger than the entire sample, and we tested it in two ways. First, from the ratio of adjacent planet parameters ($P_{i+1}/P_{i}$) distributions we selected those planet pairs whose parameter ratio is closest to unity. By removing values farthest from unity, and performing the Pearson test at every step, we obtain the high-correlated (HC) subsamples of masses, radii, and densities in pairs and period ratios in triples with Pearson R $\sim$ 0.95 (HCR, HCM, HCD, and HCP subsamples). We used these new subsamples for cross-checking analyses for mass and radius, and interior composition similarity for planets belonging to the same subsample.

Second, we performed a `moving window' test, similar to \citet{otegi2021similarity}: (i) we worked with separated subsamples for masses, radii, densities and period ratios of adjacent planets; (ii) we sorted all planet pairs in each subsample ascending for a given stellar parameter, then selected the first 40 data points and perform the Pearson test; (iii) we repeated the same procedure with a continuously moving 40-pair subsample window until the entire subsample is covered. This technique allows us to identify the ranges in stellar parameters, where subsamples show an unusual correlation. The significance of this tendency was assessed by a bootstrap test: the correlation coefficients of the partial sample were compared with a median value of those of distributions from 10$^5$ simulations, in which we calculated the Pearson coefficients for a random subset of the same size drawn from the entire sample. The bootstrap technique is typically used to assess non-normal data distributions, as in this study, and we used this bootstrap experiment as an additional test for the previously obtained Pearson correlation coefficients for mass, radius, density and period ratio subsamples.

\section{Results}
\label{sec:results}
\subsection{Mass, radius, density, and period ratio analyses}

\begin{figure*}[ht!]\resizebox{\hsize}{!}{
   \includegraphics[width=18cm]{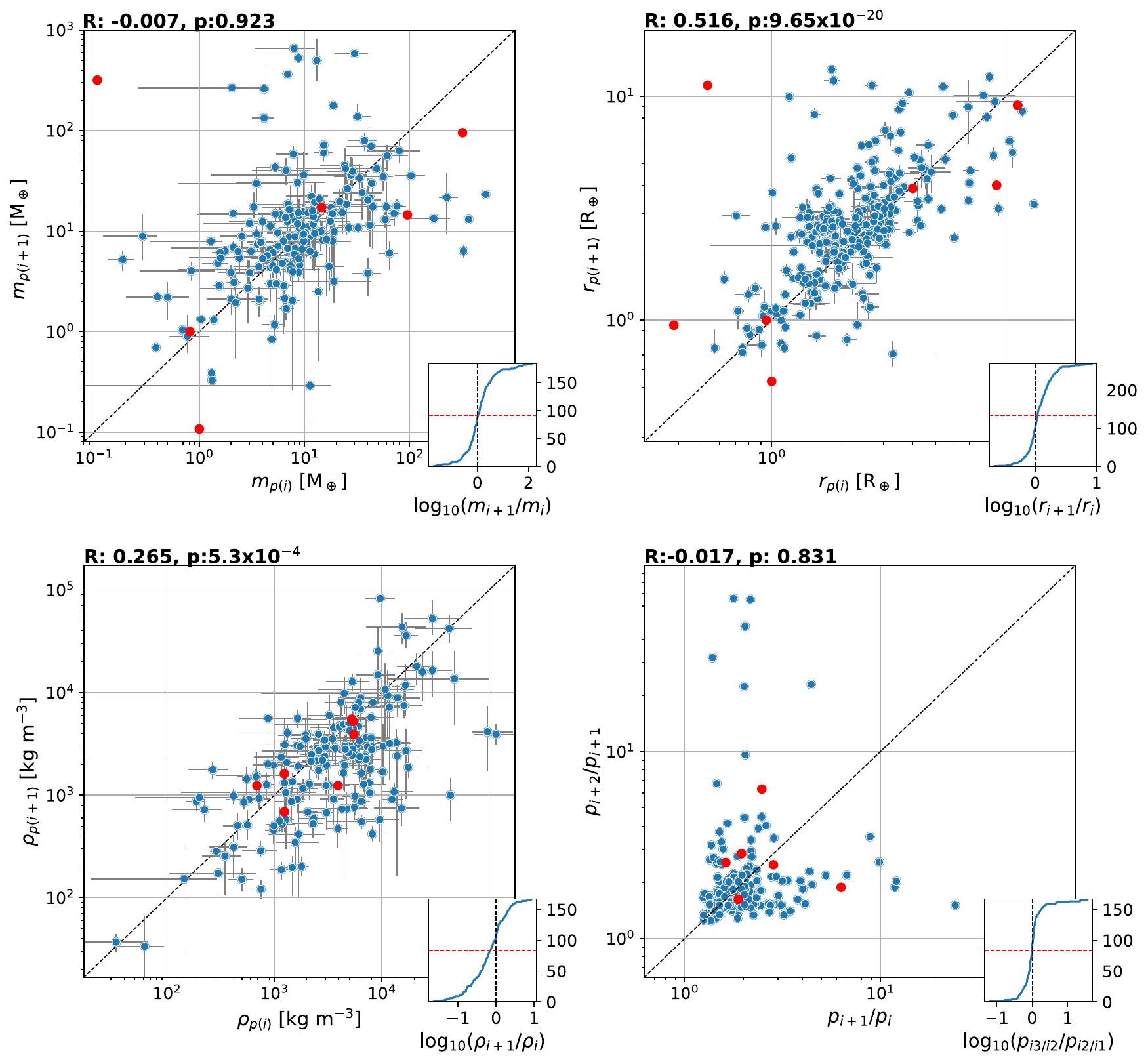}
}
    \caption{Masses, radii, densities, and period ratios of adjacent planets. The scatter graph for mass (top left), radius (top right), and density (bottom left) of a planet against the same parameter of the next planet farther from the star is plotted with observational errors indicated by the grey lines. The bottom right panel shows the period ratio between the middle and inner planet against that between the outer and middle planet in triples. The blue and red dot-shaped markers correspond to the sample and Solar System planet pairs, respectively. The black dashed line is the 1:1 line. The observational errors could be smaller than the marker size. In the bottom right corner of each plot, the cumulative distribution function is plotted for the decimal logarithm of each planet parameter ratio (x-axis) versus counts (y-axis). In these subplots, the black vertical dashed line is unity and the red horizontal dashed line is where the cumulative distribution function reaches the 50th percentile.}\label{fig:1}
\label{fig:1}
\end{figure*}

We analysed the peas in a pod trend in our sample. First, we tested whether the pairs of adjacent planets in our sample have similar mass, radius and density, or whether the planet triples show similarity in period ratios, by using the Pearson correlation test. Second, we assessed intrasystem similarity, calculating the parameter dispersion for each individual system and the mean value for the entire sample. As we aimed to assess similarity patterns in the architecture of planetary systems in the general interpretation formulated by \citet{mishra2021new}, we needed to first evaluate the sample of all multi-planet systems with known fundamental parameters for at least two planets, and then test our findings by posing multiple restrictions on the sample, including the error truncation.  Although it seems more practical only to study planets with well-determined (>3$\sigma$) masses, imposing such a significance criterion biases the sample toward more massive planets at a given radius. This is why we suggested the error-accommodation routine that "weights" the data points according to the measurement precision to fully accommodate the error of the different parameters. The routine allowed us to consolidate a sample large enough to represent multi-planet systems statistics. Our sample consists of data for 306 planets in 122 systems with determined masses, 414 planets in 145 systems with determined radii, and 353 planets in 93 systems with determined periods.

\subsubsection{Correlation in the  parameters of adjacent planets}
In Fig.\ref{fig:1}, the scatter graphs represent the inner planet parameter plotted against the outer planet parameter: mass, radius and density, as well as period ratio between the middle and inner planet plotted against period ratio between the outer and middle planet in triples. The outer planets with larger radius comprise 62\% of the sample, and this "ordered in size" trend was previously suggested by \citet{weiss2018california}. 52.7\% of the mass pairs are above the unity line, indicating that there is no obvious trend. Only 35.3\% of the density pairs are above the 1:1 line, which means that more distant planets tend to be less dense. For period ratios, spacing in triples lies almost equally above (50.3\%) and below (49.7\%) the unity line.

We find that there is a moderate correlation for radii of adjacent planets with the Pearson R=0.516 and p-value $\sim$10$^{-19}$, trumpeting R as significant, and a weak correlation for densities R=0.265 with p = 0.0005, which is less than adopted $\alpha$=0.05. There is no correlation for masses R=-0.007 or period ratios R=-0.017, and the 95\% confidence intervals for these parameters do not give grounds to assume even a weak correlation for the planets in the main sample. The results
for the sample and after running 10$^5$ simulations using the error-accommodating technique (see Sect.\ref{sec:methods}) do not differ much and are represented in Table \ref{table:1} and in Fig.\ref{fig:23a} in Appendix \ref{sec:appendix}. We repeated the analysis with a restricted subsample with higher parameter precision of $\sigma_{P\mathrm{sum}}\leq$ 0.5$P$, and it showed similar results, except for the density subsample, which revealed a moderate correlation with R=0.654 in the small subsample (n=44). We created 10$^5$ bootstrapped samples with 40 random pairs from each parameter subsample and subsequently performed correlation tests. The bootstrap test results confirmed our initial outcome, and along with the error precision test results are also represented in Table \ref{table:1}. and additionally in Fig.\ref{fig:29} in Appendix \ref{sec:appendix}. As the last test, we constructed 10$^5$ mass and density test populations using the non-parametric model \citep{2018ApJ...869....5N}  for the observed radii in our sample and repeated the calculation of the Pearson coefficients. For mass and density, the expected values in the main sample are within 1$\sigma$, and for the error precision of $\sigma_{P\mathrm{sum}}\leq$ 0.5$P$ we found that the expected value for density correlation is larger than for the main sample (see Table \ref{table:1}).

To examine the robustness of our results concerning the similarity of adjacent planets in the sample, we conducted several tests along with those already mentioned. Our original sample consists of data for 246 planet pairs from the Kepler survey and 47 pairs of planets discovered by TESS and the rest discovered by others, and therefore the sample could suffer from instrument-related biases. To eliminate this effect, we analysed the data from Kepler separately and listed results in Table \ref{table:1}). The Kepler subsample shows similar results as the main sample in the Pearson test. Then, we tested the sensitivity of the results on the sample size and parameter error precision by restricting the entire sample to $\sigma_{P\mathrm{sum}}/P \leq 1/2$. This test clarifies that the results do not deviate much with increasing data accuracy, except for densities, which show a moderate correlation in the restricted sample. Moreover, the test population produced from the non-parametric MRR shows analogous results in mass and density for the Kepler sample.

Finally, we tested the sample restrictions often found in the literature: M$_p$ < 100 M$_\oplus$ and R$_p$ < 10 R$_\oplus$ (see e.q. \citet{Wang_2017,otegi2021similarity}). This restricted subsample consists of 73 pairs of planets with error precision adopted in this study and the Pearson coefficients show a medium to strong correlation in radius (R = 0.69), a weak to medium correlation in mass (R = 0.35) and very weak correlation in density (R = 0.16).

\subsubsection{Intrasystem dispersion}
For the entire sample of systems with at least two planets, we found that the average intrasystem dispersion $D$ in radius is 0.317, and it is larger for mass ($D$=0.54), for density ($D$=0.584), which indicates that radii of planets belong to one system are more uniform than masses and densities. We run 10$^5$ simulations using the error-accommodating test for these parameters, and the results show a slightly larger dissimilarity in mass and density. Using the MRR from the non-parametric model \citep{2018ApJ...869....5N} we simulated 10$^5$ mass and density test populations for the observed radii in our sample and repeated the calculation of $D$. These results show that the expected values of $D$ point towards an even larger dispersion than the error accommodation test, but it is still less than the simulated mean for mass and density (see Fig \ref{fig:28c}, two left plots, in Appendix \ref{sec:appendix}). The repeated analysis with a restricted subsample with the parameter precision of $\sigma_{P\mathrm{sum}}\leq$ 0.5$P$ showed similar results.

From the main sample, we selected systems with at least 3 planets and repeated the routine, adding calculations for the period ratio between two consecutive pairs of planets in the system. Again, the radius is the only parameter that is consistently more uniform than the other parameters we investigated. Remarkably, the systems with at least three planets show less uniformity in mass and density, than the entire sample (D$\approx$0.54-0.58 for N$_{pps}$>1 and D$\approx$0.71-0.74 for N$_{pps}$>2, where N$_{pps}$ is the number of planets per system). It could be the indication, that 2-planet systems, which is a majority in our sample, are more uniform than systems with higher multiplicity, but we noted that such systems can be incomplete, and we discuss the completeness of systems in our sample in Section \ref{sec:discussion}. It also confirms our conclusion based on the Pearson correlation test and shows that the radii tend to be similar both in pairs of adjacent planets and within the system they belong to. However other parameters show more dissimilarity in both cases. The calculation of $D$ for the main sample and subsamples is represented in Table \ref{table:15}.

\begin{figure}[ht!]
\resizebox{\hsize}{!}{

\includegraphics{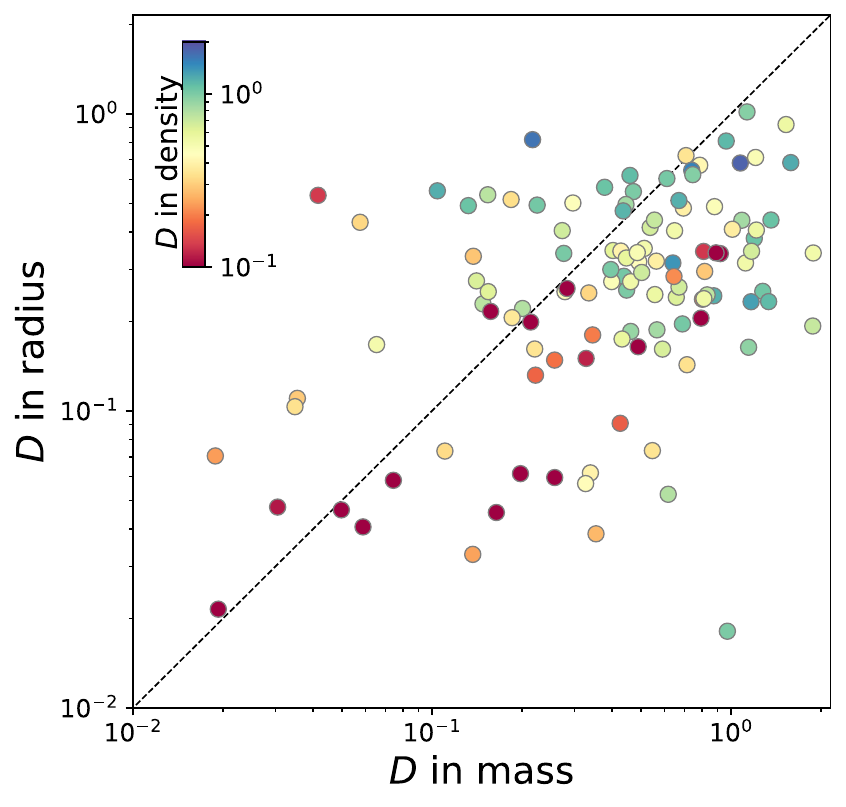}
}
\caption{Intrasystem dispersion in mass, radius, and density. The x-axis shows dispersion in mass for systems in the main sample, the y-axis represents dispersion in radii, and the colour of the dots corresponds to dispersion in density. The red colours indicate similarity in density, and the blue colours indicate dissimilarity. The black dashed line is 1:1.}
\label{fig:41}
\end{figure}

In Fig.\ref{fig:41}, the scatter graphs represent all systems in our sample, for which intrasystem dispersion in mass, radius and density was calculated.
The number of systems which have low dispersion in radius is larger than that in mass. Low dispersion in density appeared in systems with large dispersion in mass. Among systems with both small dispersion in mass and radius, there are systems with large dispersion in density, and in systems with dissimilar mass or radius, density can be similar. More systems are below the unity line, which not only indicates that more systems are similar in radius than in mass, but the systems with similar planet masses differ from those hosting planets with similar radii.

Usually, the significance of the intrasystem uniformity $D$ is assessed by comparing its value to a Monte Carlo distribution of values generated by randomly shuffling planets among the set of systems or drawing one planet parameter randomly from the observed population (e.g., \citet{millholland2017kepler,Wang_2017,otegi2021similarity,2023MNRAS.525L..66L}). The former approach can lead to instability in the modelled systems due to the adjacent pair of mock planets ending up too close to each other. The latter approach could produce non-physical systems of planets with an unnatural set of fundamental parameters unlike any real system, and the drawback of this was described by \citet{millholland2021split} as such: the erasing of the structure in period and radius distributions would lead to an overestimate of the strength of intrasystem uniformity. We therefore chose another method of producing test populations which is necessary for the null hypothesis testing. Instead, for mass and density, we produced multiple (10$^5$) test populations using the non-parametric MRR \citep{2018ApJ...869....5N} and observed radii in our sample and subsamples, retaining only planets that satisfied dynamical stability requirements (see Section \ref{sec:methods}). We tested intrasystem similarity by measuring the dispersion in fundamental parameters in our sample, including only the systems with planets M$_{pl}$ < 100 M$_\oplus$. In addition to the main and mass-restricted sample, we also computed intrasystem dispersions in the TTV and RV subsamples separately. For masses, densities and period ratios the values of $D$ in all cases were consistently larger than 0.5 with the exception of $D$ in mass in restricted M$_{pl}$ < 100 M$_\oplus$ subsample with $D$= 0.476$\pm$0.013. In this study, for most subsamples and the entire sample, $D$ in mass is typically larger than 0.5, whereas our calculations show that dispersion in density is typically larger than 0.6 and exceeds the $D$ value for mass and radius.

\begin{table*}
    \caption{\label{table:1} Pearson’s R and p-value for the radii (r), masses (m), densities ($\rho$), and period ratios (pr) of adjacent planets}
    \centering
    \small
    \begin{tabularx}{17cm}{c c c c c c c c c}

    \multicolumn{9}{c}{\textbf{Main sample correlation}}\TBstrut\\
    \hline
    \hline
     & N$_P$, N$_s$, N$_{pl}$ & p-value & R  & CI & p$_{err}$ & $\widetilde{X}$(R)&CI$_{err}$&$\overline{R_{sim}}$ \TBstrut\\
    \hline
     $\sum\frac{(\sigma_{r\pm})}{r} \leq$ 2 & 269, 145, 414 & 9.7$\times$10$^{-20}$ & 0.516 & [0.423; 0.599] & 1.28$\times$10$^{-19}$ & 0.511& [0.417; 0.595]&-\TBstrut\\
     $\sum\frac{(\sigma_{m\pm})}{m} \leq$ 2 & 184, 122, 306 & 0.92 & -0.007 & [-0.152; 0.138] & 0.538 & -0.007 &[-0.152; 0.138]& 0.095$\pm$0.130\TBstrut\\
     $\sum\frac{(\sigma_{\rho\pm})}{\rho}  \leq$ 2& 167, 115, 282 & 0.0005 & 0.265 & [0.118; 0.401] & 0.0004 & 0.251&[0.109; 0.393]& 0.225$\pm$0.108\TBstrut\\
     $\sum\frac{(\sigma_{pr\pm})}{pr}  \leq$ 2 & 167, 93, 353 & 0.83 & -0.0166 & [-0.168; 0.136] & 0.584 & -0.017 & [-0.168; 0.136]&- \TBstrut\\
    \hline
    \\
    \multicolumn{9}{c}{\textbf{Error precision test}}\TBstrut\\
    \hline
    \hline
     $\sum\frac{(\sigma_{r\pm})}{r} \leq$ 0.5 & 263, 141, 404 & 1.2$\times$10$^{-15}$ & 0.467 & [0.367; 0.556] & 1.1$\times$10$^{-15}$ & 0.463 & [0.362; 0.553]&-\TBstrut\\
     $\sum\frac{(\sigma_{m\pm})}{m} \leq$ 0.5 & 70, 52, 123 & 0.83 & -0.027 &[-0.260; 0.210]& 0.581 & -0.025 & [-0.258; 0.211] &0.048$\pm$0.133\TBstrut\\
     $\sum\frac{(\sigma_{\rho\pm})}{\rho}  \leq$ 0.5 & 44, 33, 77 & 1.4$\times$10$^{-6}$  & 0.654 &[0.443; 0.796] & 1.2$\times$10$^{-6}$  & 0.644&[0.430; 0.790] &0.385$\pm$0.163\TBstrut\\
     $\sum\frac{(\sigma_{pr\pm})}{pr}  \leq$ 0.5 & 165, 92, 349 & 0.99& -0.001 & [-0.154; 0.152] & 0.99 & -0.001& [-0.154; 0.152] &-\TBstrut\\
    \hline
     $\sum\frac{(\sigma_{r\pm})}{r} \leq$ 0.32 & 239, 136, 378 & 4.4$\times$10$^{-15}$ & 0.479 & [0.374; 0.571] & 3.7$\times$10$^{-15}$ & 0.475& [0.371; 0.568]&-\TBstrut\\
    $\sum\frac{(\sigma_{m\pm})}{m} \leq$ 1 & 136, 92, 229 & 0.98 & 0.002 & [-0.167; 0.170] & 0.496 & 0.001 &[-0.168; 0.169]& 0.089$\pm$0.134\TBstrut\\
     \hline
    \\
    \multicolumn{9}{c}{\textbf{Kepler planets subsample correlation}}\TBstrut\\
    \hline
    \hline
     $\sum\frac{(\sigma_{r\pm})}{r} \leq$ 2 & 229, 120, 349 & 2.1$\times$10$^{-15}$ & 0.493 & [0.388; 0.585] & 2.7$\times$10$^{-15}$ & 0.486& [0.381; 0.580]&-\TBstrut\\
     $\sum\frac{(\sigma_{m\pm})}{m} \leq$ 2 & 145, 97, 242 & 0.75 & -0.026 &[-0.189; 0.137]& 0.623 & -0.027&[-0.189; 0.137] &0.083$\pm$0.131\TBstrut\\
     $\sum\frac{(\sigma_{\rho\pm})}{\rho}  \leq$ 2& 145, 91, 220 & 0.004 & 0.253 &[0.084; 0.408]& 0.002 & 0.239&[0.076; 0.401] &0.230$\pm$0.116\TBstrut\\
     $\sum\frac{(\sigma_{pr\pm})}{pr}  \leq$ 2 & 143, 80, 303 & 0.98 & -0.002 & [-0.166; 0.162]&0.51 & -0.006& [-0.166; 0.162]&- \TBstrut\\
    \hline
    \end{tabularx}
    \begin{tabularx}{17cm}{c c c c c c c c c c c c c c}
     \\
    \multicolumn{14}{c}{\textbf{Radii from Gaia DR3 correlation}}\TBstrut\\
    \hline
    \hline
    &    N$_P$, N$_s$, N$_{pl}$  &&  p$_\mathrm{flame}$ &&  R$_\mathrm{flame}$ && CI$_\mathrm{flame}$ && p$_\mathrm{gspphot}$ && R$_\mathrm{gspphot}$  && CI$_\mathrm{gspphot}$ \TBstrut\\
    \hline
     $\sum\frac{(\sigma_{r\pm})}{r} \leq$ 2 & 227, 126, 353 && 5.9$\times$10$^{-13}$ && 0.454 && [0.344;  0.552] &&1.4$\times$10$^{-12}$ && 0.447 && [0.337;  0.546]\TBstrut\\
     $\sum\frac{(\sigma_{r\pm})}{r} \leq$ 0.5 & 223, 124, 347 && 2.7$\times$10$^{-11}$ && 0.427 && [0.313; 0.529] &&6.0$\times$10$^{-11}$ && 0.420 && [0.306;  0.523]\TBstrut\\
    \hline
    \end{tabularx}
    \begin{tabularx}{17cm}{c c c c c c c c c c c c c c c}
    \\
     \multicolumn{15}{c}{\textbf{HCR, HCM, HCD, HCP subsamples}}\TBstrut\\
    \hline
    \hline
         &&N$_P$, N$_s$, N$_{pl}$& & p-value && R  && CI && p$_{err}$-value && $\widetilde{X}$(R)&&CI$_{err}$ \TBstrut\\
         \hline
     $\sum\frac{(\sigma_{r\pm})}{r} \leq$ 2 && 149, 97, 254& & 5.7$\times$10$^{-82}$ && 0.958 && [0.943; 0.970] &&5.86$\times$10$^{-70}$&& 0.943&& [0.920; 0.958]\TBstrut\\
     $\sum\frac{(\sigma_{m\pm})}{m} \leq$ 2 && 67, 61, 130 && 1.7$\times$10$^{-37}$ && 0.959 && [0.935 0.975]&& 1.2$\times$10$^{-24}$ && 0.905&&[0.840; 0.938]   \TBstrut\\
     $\sum\frac{(\sigma_{\rho\pm})}{\rho}  \leq$ 2&& 69, 59, 129 && 2.6$\times$10$^{-38}$ && 0.959 &&[0.954; 0.984] && 4.5$\times$10$^{-7}$ && 0.904  & &[0.821; 0.933]  \TBstrut\\
     $\sum\frac{(\sigma_{pr\pm})}{pr}  \leq$ 2 &&43, 34, 112 && 1.2$\times$10$^{-21}$ && 0.956 && [0.920; 0.976]&& 4.0$\times$10$^{-14}$&&0.966 && [0.920; 0.986]\TBstrut\\
    \hline
    \end{tabularx}
    \begin{tabularx}{17cm}{c c c c c c c c c c c}
    \\
    \multicolumn{11}{c}{\textbf{HCR subsample correlation}}\TBstrut\\
    \hline
    \hline
    && N$_P$, N$_s$, N$_{pl}$ & p-value & R  & CI & p$_{err}$ &$\widetilde{X}$(R) & CI$_{err}$ &p$_{sim}$ & $\overline{R_{sim}}$\TBstrut\\
    \hline
    $\sum\frac{(\sigma_{m\pm})}{m} \leq$ 2 && 91, 74, 168 & 0.0016 & 0.326&[0.130; 0.499]&0.0007&0.333&[0.135; 0.503]&0.16&0.355$\pm$0.267 \TBstrut\\
    $\sum\frac{(\sigma_{\rho\pm})}{\rho} \leq$ 2 && 84, 67, 154 & 0.00001 & 0.463 &[0.275; 0.616]&3.3$\times$10$^{-5}$&0.415 &[0.228; 0.583]&0.012&0.443$\pm$0.142\TBstrut\\
    \hline
    \end{tabularx}
    \begin{tabularx}{17cm}{c c c c c | c c c c c}
    \\
     \multicolumn{5}{c}{\textbf{Stellar HCM and HCD subsamples (N$_P$=40)}}&\multicolumn{4}{c}{\textbf{Bootstrap test (N$_P$=40)}} \TBstrut\\
    \hline
    \hline
     subsample type& N$_s$, N$_{pl}$ & p-value & R & CI & & p$_{bts}$& $\widetilde{X}$(R)$_{bts}$ & CI$_{bts}$ \TBstrut\\
         \hline
     m T$_\mathrm{eff}$ [2566; 4398] K &  24, 64& 5.6$\times$10$^{-15}$ & 0.896& [0.811; 0.944]&$\sum\frac{(\sigma_{r\pm})}{r}$ $\leq$ 2 & 1.7$\times$10$^{-22}$ &   0.545&[0.455; 0.624]  \TBstrut\\

     $\rho$ [-0.21; -0.01] [Fe/H]& 31, 71& 3.99$\times$10$^{-14}$ & 0.884 &[0.790; 0.937]&$\sum\frac{(\sigma_{m\pm})}{m}$ $\leq$ 2 & 0.5 &   0.002&[-0.143; 0.146]  \TBstrut\\

     $\rho$ [6.03; 12.3] Gyr &24, 64 & 3.7$\times$10$^{-12}$ & 0.850  &[0.733  0.919]& $\sum\frac{(\sigma_{\rho\pm})}{\rho}$ $\leq$ 2& 4.5$\times$10$^{-7}$ &   0.362& [0.223; 0.487]\TBstrut\\
      &&  &  && $\sum\frac{(\sigma_{pr\pm})}{pr}$ $\leq$ 2 & 0.21 &   0.066&[-0.097;  0.226]\TBstrut\\
    \hline
    \end{tabularx}
    \tablefoot{N$_P$, N$_s$, N$_{pl}$ stand for the number of pairs/triples, the total number of systems, and the total number of planets in the subsample, respectively. $\widetilde{X}$(R) and $\overline{R_{sim}}$ stand for median and mean R-value, respectively, p is the p-value, and CI is the confidence interval stated for $\alpha$ = 0.05. Prefix $err$ stands for the results of the error-accommodation test for uncertainties. Prefix $sim$ stands for results after 10$^5$ simulations of mass and density populations from the non-parametric MRR \citep{2018ApJ...869....5N} giving the observed radii. Prefix $bts$ stands for results after the bootstrap test was performed on the entire sample.
    }
\end{table*}
\begin{table}[ht!]
    \caption{\label{table:15}Intra-system dispersion $D$ for radii (r), masses (m), densities ($\rho$) and period ratios (p)}
    \centering
    \small
    \begin{tabularx}{8,5cm}{c c c c c }%
    \\
    \multicolumn{5}{c}{\textbf{Main sample $D$, N$_{pps}$ >1, $\sum\frac{(\sigma_{p\pm})}{p} \leq$ 2}}\TBstrut\\
    \hline
    \hline
      & N$_{pl}$, N$_s$ & $D$ & $\overline{D_{err}}$ & $\overline{D_{sim}}$\TBstrut\\
    \hline
    r &  493, 171 & 0.317 & 0.322$\pm$0.003 & -\TBstrut\\
     m &  319, 127 & 0.540 & 0.566$\pm$0.012 & 0.647$\pm$0.031 \TBstrut\\
     $\rho$ & 296, 120 & 0.584 & 0.619$\pm$0.047 & 0.706$\pm$0.031 \TBstrut\\
     \hline
    \\
    \multicolumn{5}{c}{\textbf{$D$, N$_{pps}$ >1, $\sum\frac{(\sigma_{r\pm})}{r} \leq$ 0.5}}\TBstrut\\
    \hline
    \hline
    r & 483, 167 & 0.318 & 0.320$\pm$0.003& -\TBstrut\\
     m &  137, 58 & 0.494 & 0.500$\pm$0.008 & 0.647$\pm$0.057 \TBstrut\\
     $\rho$ & 86, 36 & 0.500 & 0.505$\pm$0.013 & 0.714$\pm$0.06 \TBstrut\\
     \hline
     \\
    \multicolumn{5}{c}{\textbf{$D$ for subsample N$_{pps}$ >2}}\TBstrut\\
    \hline
    \hline
     r& 329, 89 & 0.341 & 0.343$\pm$0.003& - \TBstrut\\
     m & 147, 41 & 0.711 & 0.727$\pm$0.014 & 0.761$\pm$0.051   \TBstrut\\
     $\rho$  & 130, 37 & 0.742 & 0.762$\pm$0.020 & 0.862$\pm$0.053  \TBstrut\\
     pr& 374, 99 & 0.615 & 0.615$\pm$0.002 & -\TBstrut\\
    \hline
    \\
    \multicolumn{5}{c}{\textbf{$D$ for subsample M$_p$ < 100 M$_\oplus$}}\TBstrut\\
    \hline
    \hline
     r & 422, 146 & 0.284 & 0.289$\pm$0.003&- \TBstrut\\
     m & 282, 113 & 0.446 & 0.476$\pm$0.013 & 0.615$\pm$0.032\TBstrut\\
    $\rho$ & 266, 107 & 0.550 & 0.588$\pm$0.052 & 0.699$\pm$0.033 \TBstrut\\
    pr (N$_{pps}$ >2)& 317, 84 & 0.559 & 0.558$\pm$0.002&- \TBstrut\\
    \hline
    \\
    \multicolumn{5}{c}{\textbf{$D$ for RV subsample}}\TBstrut\\
    \hline
    \hline
     r & 171, 66 & 0.337 & 0.340$\pm$0.005&- \TBstrut\\
     m & 140, 55 & 0.502 & 0.521$\pm$0.016 & 0.645$\pm$0.046\TBstrut\\
     $\rho$& 132, 53 & 0.561 & 0.581$\pm$0.308 & 0.722$\pm$0.048\TBstrut\\
     pr (N$_{pps}$ >2) &107, 29& 0.689 & 0.689$\pm$0.000&- \TBstrut\\
    \hline
    \\
    \multicolumn{5}{c}{\textbf{$D$ for TTVs subsample}}\TBstrut\\
    \hline
    \hline
     r & 166, 71 & 0.230 & 0.238$\pm$0.005 & -\TBstrut\\
    m &  156, 65 & 0.517 & 0.547$\pm$0.018 & 0.642$\pm$0.044 \TBstrut\\
     $\rho$ & 146, 61 & 0.585 & 0.631$\pm$0.022 & 0.679$\pm$0.045 \TBstrut\\
     pr (N$_{pps}$ >2) &61, 17& 0.455 & 0.455$\pm$0.000&- \TBstrut\\
    \hline
    \\
    \multicolumn{5}{c}{\textbf{$D$ for HCR subsample}}\TBstrut\\
    \hline
    \hline
     r& 250, 96 & 0.108 & 0.116$\pm$0.004& - \TBstrut\\
     m & 172, 75 & 0.351 & 0.388$\pm$0.016 & 0.561$\pm$0.040 \TBstrut\\
     $\rho$ & 158, 68 & 0.398 & 0.437$\pm$0.020 & 0.579$\pm$0.041  \TBstrut\\
    \hline
    \\
    \multicolumn{5}{c}{\textbf{$D$ for radius from Gaia DR3}}\TBstrut\\
    \hline
    \hline
    r$_\mathrm{flame}$ &  419, 147 & 0.333 & 0.333$\pm$0.002& -\TBstrut\\
    r$_\mathrm{gspphot}$ & 419, 147 & 0.333 & 0.335$\pm$0.003 &-\TBstrut\\
    \hline
    \end{tabularx}
    \tablefoot{N$_{pps}$ stand for the number of planets per system, and N$_s$, N$_{pl}$ the same as in Table \ref{table:1}. $\overline{D_{err}}$ stands for average dispersion and $\sigma$ in parameter after the error-accommodation test for uncertainty to our exoplanet sample. $\overline{D_{sim}}$ stands for average dispersion and $\sigma$ after 10$^5$ simulation of mass and density populations from the non-parametric MRR \citep{2018ApJ...869....5N} giving the observed radii.}
\end{table}
\begin{table}[ht!]
    \caption{\label{table:6}The interior distance $I$ in pairs of planets for HCR subsample}
    \centering
    \small
\begin{tabular}{l c c c c}
    \hline
    \hline
     & $I$ & $\overline{I_{err}}$ & $\overline{I_{sim}}$ & $\widetilde{X}$(I$_{sim}$)\TBstrut \\
     \hline
Kepler-32 b-c & 0.010 & 0.119$\pm$0.076& 0.049$\pm$0.012 & 0.044\TBstrut \\
TRAPPIST-1 e-f & 0.010 & 0.018$\pm$0.011& 0.088$\pm$0.013 & 0.087\TBstrut \\
TRAPPIST-1 b-c & 0.014 & 0.019$\pm$0.011& 0.015$\pm$0.010 & 0.012\TBstrut \\
TRAPPIST-1 f-g & 0.017 & 0.020$\pm$0.011& 0.026$\pm$0.016 & 0.024\TBstrut \\
TOI-561 d-e & 0.019 & 0.051$\pm$0.033& 0.056$\pm$0.016 & 0.049\TBstrut \\
TOI-776 b-c & 0.023 & 0.090$\pm$0.061& 0.069$\pm$0.010 & 0.065\TBstrut \\
K2-138 c-d & 0.026 & 0.088$\pm$0.051& 0.075$\pm$0.013 & 0.070\TBstrut \\
TRAPPIST-1 d-e & 0.026 & 0.019$\pm$0.011& 0.144$\pm$0.015 & 0.144\TBstrut \\
HD 219134 b-c & 0.037 & 0.045$\pm$0.027& 0.053$\pm$0.007 & 0.050\TBstrut \\
Kepler-176 c-d & 0.039 & 0.147$\pm$0.089& 0.109$\pm$0.012 & 0.10\TBstrut \\
TOI-270 c-d & 0.042 & 0.058$\pm$0.027& 0.085$\pm$0.013 & 0.081\TBstrut \\
Kepler-1705 b-c & 0.049 & 0.068$\pm$0.045& 0.015$\pm$0.014 & 0.009\TBstrut \\
LTT 1445 A c-b & 0.053 & 0.060$\pm$0.040& 0.043$\pm$0.012 & 0.042\TBstrut \\
K2-3 b-c & 0.082 & 0.147$\pm$0.100& 0.100$\pm$0.015 & 0.099\TBstrut \\
K2-146 b-c & 0.097 & 0.171$\pm$0.081& 0.125$\pm$0.012 & 0.121\TBstrut \\
Kepler-28 b-c & 0.109 & 0.136$\pm$0.083& 0.038$\pm$0.011 & 0.034\TBstrut \\
Kepler-161 b-c & 0.113 & 0.177$\pm$0.114& 0.080$\pm$0.012 & 0.076\TBstrut \\
TOI-1260 b-c & 0.115 & 0.127$\pm$0.071&0.172$\pm$0.013 & 0.167\TBstrut \\
HD 260655 b-c & 0.117 & 0.117$\pm$0.048& 0.134$\pm$0.016 & 0.139\TBstrut \\
HD 136352 c-d & 0.119 & 0.120$\pm$0.023& 0.126$\pm$0.015 & 0.120\TBstrut \\
K2-138 d-e & 0.126 & 0.166$\pm$0.055& 0.216$\pm$0.014 & 0.210\TBstrut \\
L 98-59 c-d & 0.133 & 0.132$\pm$0.061& 0.076$\pm$0.009 & 0.077\TBstrut \\
TOI-178 e-f & 0.149 & 0.151$\pm$0.070& 0.036$\pm$0.014 & 0.030\TBstrut \\
Kepler-128 b-c & 0.160 & 0.182$\pm$0.090& 0.078$\pm$0.011 & 0.080\TBstrut \\
HD 73583 b-c & 0.165 & 0.154$\pm$0.075& 0.148$\pm$0.013 & 0.142\TBstrut \\
HD 23472 b-c & 0.165 & 0.438$\pm$0.243& 0.106$\pm$0.011 & 0.102\TBstrut \\
TOI-763 b-c & 0.172 & 0.169$\pm$0.048& 0.133$\pm$0.013 & 0.129\TBstrut \\
KOI-1599.02-01 & 0.176 & 0.176$\pm$0.095& 0.018$\pm$0.015 & 0.016\TBstrut \\
Kepler-80 d-e & 0.182 & 0.187$\pm$0.075& 0.029$\pm$0.008 & 0.025\TBstrut \\
K2-266 d-e & 0.184 & 0.202$\pm$0.094& 0.073$\pm$0.017 & 0.065\TBstrut \\
Kepler-107 b-c & 0.206 & 0.209$\pm$0.080& 0.035$\pm$0.008 & 0.032\TBstrut \\
TOI-1246 b-c & 0.207 & 0.230$\pm$0.028 & 0.175$\pm$0.015 & 0.169\TBstrut \\
K2-285 b-c & 0.210 & 0.227$\pm$0.029& 0.268$\pm$0.016 & 0.265\TBstrut \\
Kepler-138 c-d & 0.248 & 0.255$\pm$0.057& 0.106$\pm$0.012 & 0.109\TBstrut \\
Kepler-60 c-d & 0.250 & 0.270$\pm$0.085& 0.062$\pm$0.012 & 0.058\TBstrut \\
Kepler-81 b-c & 0.257 & 0.241$\pm$0.098& 0.059$\pm$0.014 & 0.054\TBstrut \\
TOI-561 f-d & 0.260 & 0.240$\pm$0.110& 0.083$\pm$0.014 & 0.078\TBstrut \\
Kepler-60 b-c & 0.272 & 0.264$\pm$0.079& 0.084$\pm$0.010 & 0.080\TBstrut \\
EPIC 220674823 b-c & 0.330 & 0.326$\pm$0.072& 0.227$\pm$0.009 & 0.225\TBstrut \\
TOI-178 f-g & 0.335 & 0.348$\pm$0.047& 0.202$\pm$0.012 & 0.197\TBstrut \\
TOI-1064 b-c & 0.362 & 0.364$\pm$0.068& 0.034$\pm$0.017 & 0.026\TBstrut \\
\hline
\end{tabular}
 \tablefoot{$\overline{I_{err}}$, $\overline{I_{sim}}$, and $\widetilde{X}$(I$_{sim}$) stand for the interior distance inferred from the error-accommodation routine, and expected mean and median values from non-parametric MRR simulations, respectively.}
\end{table}
\begin{table*}
    \caption{\label{table:2}The Pearson’s correlation coefficients (R and p-value) and mean intrasystem dispersion for period ratios (pr) subsamples}
    \centering
    \small
\begin{tabularx}{17cm}{c c c c c c c c | c c c} \\
         \multicolumn{8}{c}{\textbf{Period ratio correlation in triples}}&\multicolumn{3}{c}{\textbf{Period ratio dispersion (N$_{pps}$ >2)}} \TBstrut\\
\hline
\hline
 & N$_t$, N$_s$, N$_{pl}$ & p-value & R & CI & p$_{err}$ & $\widetilde{X}$($R_{err}$) & CI$_{err}$ &  N$_{pl}$, N$_s$& $D$&$\overline{D_{err}}$\TBstrut\\
\hline
1 & 119, 73, 267 & 0.295 & 0.097 & [-0.085,0.272] & 0.147 & 0.097 & [-0.085,0.272]&224, 62 &0.477 & 0.477$\pm$0.000\TBstrut\\
2 & 129, 76, 281 & 0.760 & -0.027 & [-0.199,0.146] & 0.620 & -0.027 & [-0.199,0.146]&309, 82& 0.593 & 0.592$\pm$0.002 \TBstrut\\
3 & 144, 83, 312 & 0.392 & 0.072 & [-0.093,0.233] & 0.196 & 0.072 & [-0.093,0.233]& 263, 73 &0.485 & 0.485$\pm$0.000 \TBstrut\\
\hline
\end{tabularx}
    \tablefoot{N$_t$ is a number of triples, N$_s$, N$_{pl}$ are the same as in Table \ref{table:1}.
    Subsample 1: r$_{pl}\le $ 6 R$_\oplus$, $pr\le $4, subsample 2: r$_{pl}\le $6, subsample 3: $pr\le $4}
\end{table*}

\subsection{Planets with similar sizes can have dissimilar masses, densities, and interior structure}

\begin{figure*}[ht!]\resizebox{\hsize}{!}{
\includegraphics{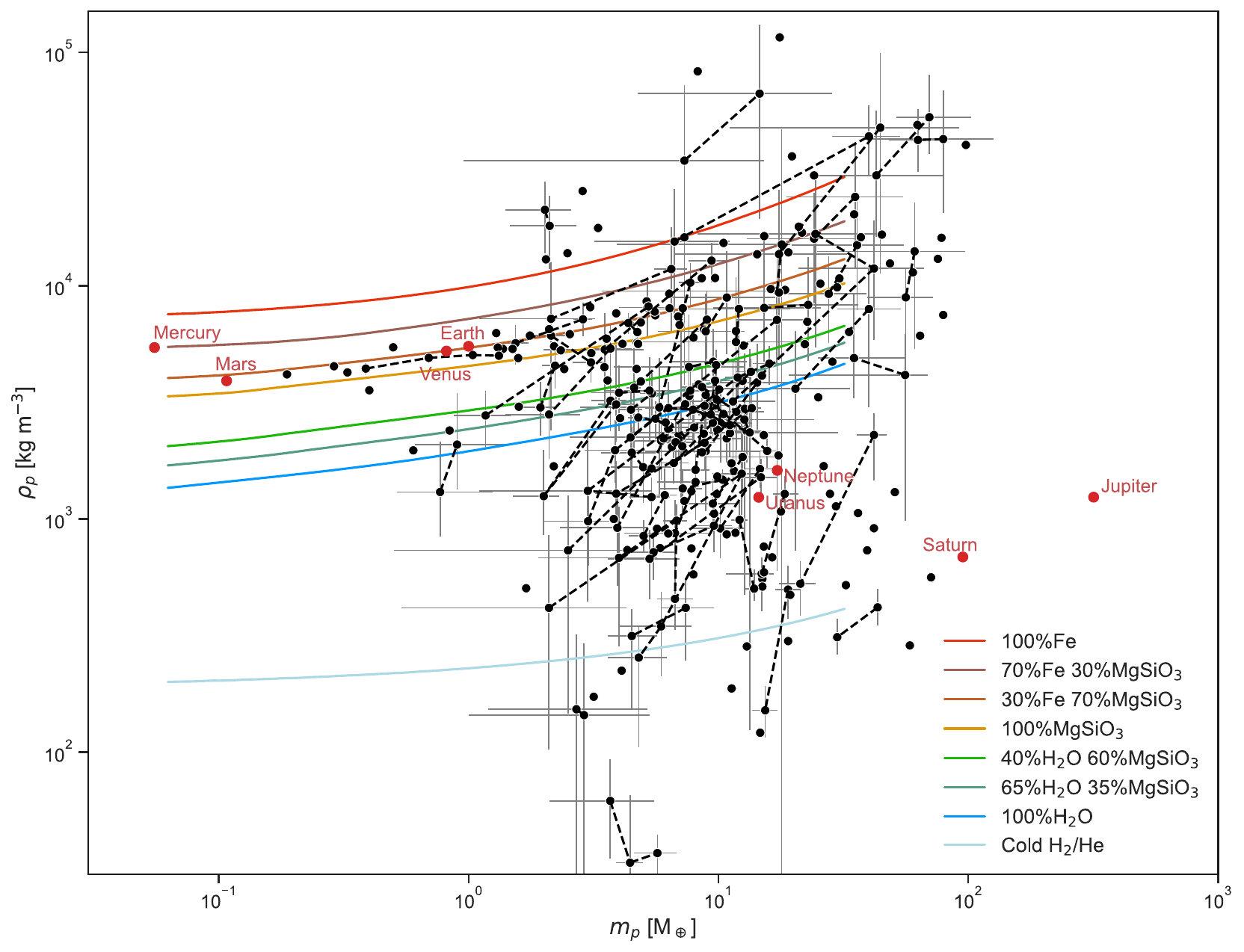}
}
\caption{The sample on the mass--density diagram. The black spheres represent the sample's planets and the red spheres are the Solar System planets, plotted for comparison. Dashed black lines connect a pair of adjacent planets in the "high correlated in radius" subsample.
The coloured interior curves (from red to light blue) are taken from \citet{zeng2016mass} and correspond to different calculated MRR for several potential 2-layer structures of exoplanets, representing different combinations of constituents of the interior: Fe, MgSiO$_3$, and H$_2$O.
The error bars are plotted for planets comprising the "high correlated in radius" subsample and omitted for others for the sake of clarity.}
\label{fig:4}
\end{figure*}

From the distributions of parameter ratios ($P_{i+1}/P_{i}$) of adjacent pairs, by sequentially removing values farthest from unity and calculating the Pearson coefficients at every step, we obtained the highly correlated subsamples of masses (HCM), radii (HCR), and densities (HCD) in pairs and period ratios (HCP) in triples with Pearson R $\sim$ 0.95 (see Table \ref{table:1} under "High correlated subsamples" section). The HCM subsample consists of only 67 pairs out of 184, which indicates that a small number of planets are similar in mass with their subsequently located neighbour. The HCR subsample, showing a high correlation in radii, consists of 149 pairs out of 269, so more than half of the sample has a small dispersion around the unity line for radius ratio. The HCD subsample consists of 69 pairs out of 167, and the HCP subsample is even smaller (43 pairs out of 167).

We proceed to study the HCR subsample further. The HCR subsample correlation in radii of adjacent planets is represented in Fig.\ref{fig:28b} in Appendix \ref{sec:appendix}, top left, along with one random realisation of bootstrapping pairs of planets from the main sample. We further investigated intrasystem dispersion in mass and density in the HCR subsample by performing the error-accommodation routine and comparing obtained $D_{err}$ with the distribution of $D_{sim}$ from 10$^5$ random simulated mass and density from non-parametric MRR, and plotted the results in Fig.\ref{fig:28c} in Appendix \ref{sec:appendix}. The observed dispersions in density and mass ($D_{err}$=0.437 and $D_{err}$=0.388, respectively) follow the same trend as the entire sample showing smaller values than the expected values $D_{sim}$, however, at the same time being larger than dispersion in radius in this subsample, which is the lowest among all samples ($D_{err}$=0.116).

In Fig.\ref{fig:4}, we plotted the entire sample in the Mass-Density diagram and connected a pair of planets from the HCR subsample with dashed lines. To guide the eye, the coloured curves in this figure correspond to several representative theoretical MRRs for 2-layer exoplanet structural models including rocky and volatile components \citep{zeng2016mass}.
Even when considering the radius and mass measurement errors, adjacent planets connected by dotted lines from the HCR subsample appear to be to a large extent well separated in their mass and bulk densities, and often straddle across distinct bulk composition curves.

To assess how strongly each parameter differs among a pair of planets, we introduce a parameter gap, $g$, to characterise the relative difference between a pair of values of a certain parameter. The term relative difference is used in quantitative science to compare two quantities while taking into account the parameter "sizes"  being compared, that is dividing by a standard, reference or starting value, or as formulated in \citet{tornqvist1985should}, an indicator of relative difference, that is a function dependent only on the ratio of parameter values, usually the arithmetic, geometric, harmonic mean, etc. The result is expressed as a ratio and is unit-less and independent of values in the sample. Therefore, the parameter gap $g$ is an absolute value of the difference between parameter values divided by the indicator of relative difference. In this case, we choose a moment mean change of second order $f(x,y) = (1/2(x^{2}+y^{2}))^{1/2}$ as an indicator of change:

\begin{equation}
    g = \frac{|x-y|}{(x^{2}+y^{2})^{1/2}},
  \label{eq:2.2}
\end{equation}
where x and y denote the values being compared. The metric $g$ varies between 0 and 1 and represents a scale-independent, normalised distance in the parameters.
At zero, the parameter gap indicates maximal similarity, and the closer this parameter is to unity, the higher is dissimilarity in the planets' pair.

To quantify the degree of mass and density correlation in the HCR subsample we first calculated the parameter gap $g$ using Eq.\ref{eq:2.2}, where $x$ and $y$ are parameters of the first and the second planet, respectively.

\begin{figure}[ht!]
\resizebox{\hsize}{!}{
\includegraphics{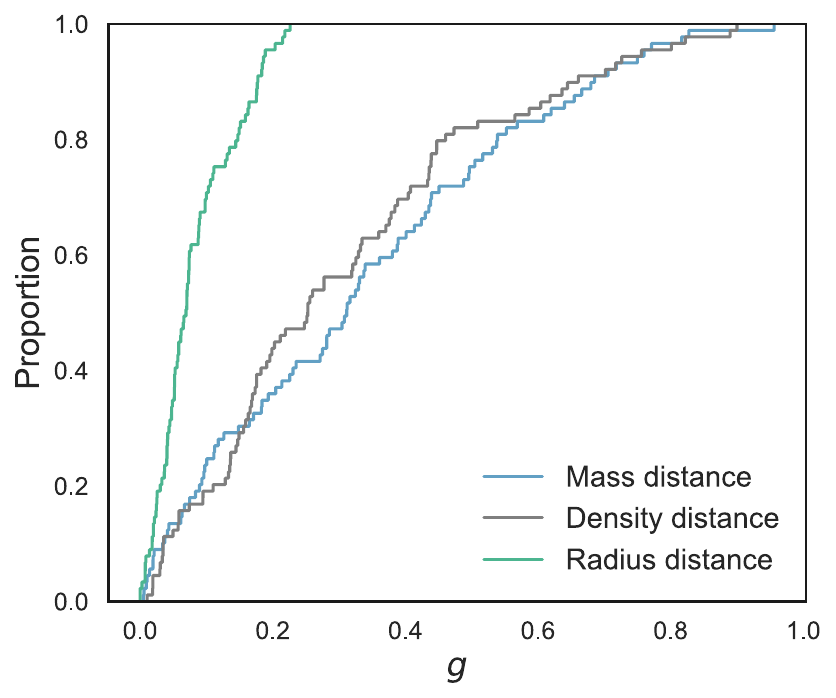}
}
\caption{ECDF of the parameter gap $g$ in mass, radius, and density in the HCR subsample. The x-axis shows $g$ in a certain parameter, the y-axis represents the proportion of the gaps in pairs that reached a certain value.}
\label{fig:51}
\end{figure}
In Fig.\ref{fig:51}, we plotted empirical cumulative distribution functions (ECDF) of $g$ in radius, mass and density of adjacent planets in the HCR subsample. ECDF represents the proportion of observations falling below each unique value in a dataset and has the advantage that each parameter $g$ is visualised directly and aids in direct comparisons between these distributions. From this figure, it is apparent that the parameter gap in radius is small for the majority of the sample. However, mass and density gaps vary in the subsample significantly, and only a third of the subsample pairs show similar gaps in all three parameters. We further probed the mass-density relationship by performing the Pearson correlation test for the HCR subsample for masses and densities of planet pairs (91 and 84, respectively) and found a weak-to-moderate correlation for both parameters instead of an expected high correlation, as for radii (see Table \ref{table:1} under "HCR subsample correlation" section). That is, in most cases, in a pair of planets from these samples, masses and bulk densities appear different, and only a very small population of adjacent planets, relative to the whole sample, will show uniformity in all three parameters.

If we place the planet pairs from the HCR sample on a fine mesh of interior structure curves in mass-density space, as shown in Fig. \ref{fig:4}, it becomes visually apparent that most pairs of planets with very similar radii straddle widely separated compositional curves, implying distinct interior structures and bulk compositions. Cases like Earth-Venus and the TRAPPIST planets, which are strung along roughly the same curves, are the exceptions.

\begin{figure}[ht!]
\resizebox{\hsize}{!}{
\includegraphics{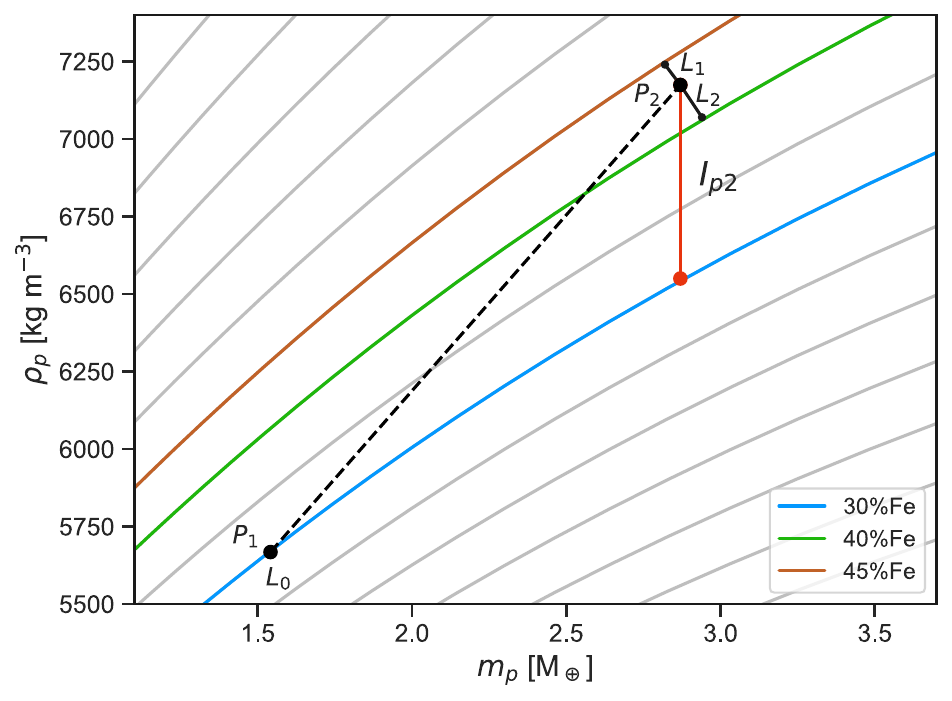}
}
\caption{
Interior difference between a pair of adjacent planets. In this illustrative example, planet $P_1$ lies closest to the blue curve (distance $L_0$) whereas planet $P_2$ is closer to the red curve (distance $L_1$). The blue curve, corresponding to 30\% Fe and 70\% MgSiO3, is chosen as the reference curve for the pair. This means if $P_2$ had the same bulk composition as $P_1$, $P_2$’s bulk density should be given by the blue curve corresponding to its mass (red circle). $I_{p1}$ and $I_{p2}$ (shown in the thin red line) are measured as the vertical gap distance between planet $P_1$ and planet $P_2$ to the reference curve at their respective masses. This figure shows LTT 1445A b and c planets, and they are separated by an interior distance I = 0.053.
}
\label{fig:82}
\end{figure}

To quantify the separation between planets in pairs in the MRR grid, we calculate the interior distance between the MRR curves that best describe each planet. For planets in our sample, that belong to the area in the Mass-Density diagram described by the aforementioned MRR curves, we first calculated the closest curve to each planet. To illustrate the concept, in Fig.\ref{fig:82} we plotted a pair of hypothetical planets with similar sizes\footnote{Based on planet parameters of LTT 1445 A c and b}. We locate the curve nearest each planet, then designate the closer of the two as the reference curve for the pair. The density values corresponding to the planet’s masses along the reference MRR curve are taken as reference densities, from which we calculate the signed difference to the actual planet densities using the gap parameter as defined in Eq.\ref{eq:2.2} (without the absolute value sign in the numerator). The mean of the two gap distances is the interior distance ($I$) for the planet pair:

\begin{equation}
    I = \frac{|I_{p1} + I_{p2}|}{2}
  \label{eq:2.3}
,\end{equation}
where $I_{pi}$ is the density gap between the planet and the reference MRR curve.

We compute $I$ for all planet pairs with at least one planet within the area covered by the grid of two-layer interiors models from \citet{zeng2016mass}, which are valid for planets with M$_{pl}\le$ 32M$_\oplus$. As a result, we excluded giant planets, as well as pairs where both planets have $I_{pi} \le$ 0.1. This value was chosen because if both planets are farther than 0.1 from any MRR curve in our set, there is no particular curve which can represent the interior composition of any planet in the pair. In this case, the resulting average $I$ would be large and would not indicate a difference between planets, but rather their mutual separation from the nearest interior curve.

\begin{figure*}
\resizebox{\hsize}{!}{\includegraphics{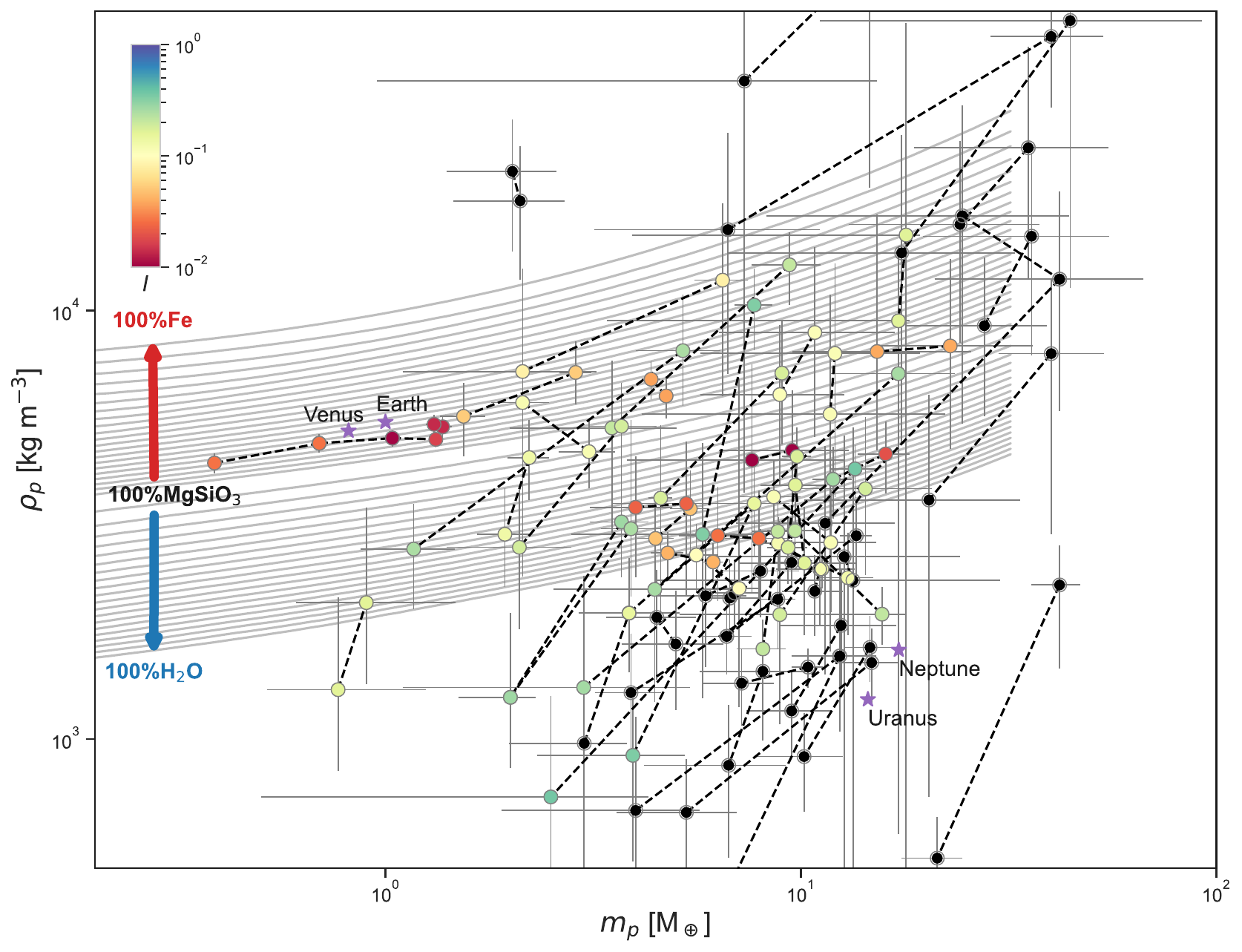}}
\caption{HCR sample on the mass--density diagram. The dots, connected by black dashed lines, are planet pairs that belong to this sample. The coloured dots represent those planets, for which we calculated interior distance $I$. The red colours indicate a pair of planets at a short interior distance, which means they are probably similar in the interior structure, and blue colours indicate planets that show larger separation in the interior from their neighbour. Light-grey curves represent the MRR interior curves calculated by \citet{zeng2016mass}. From the baseline of a silicate planet with 100\% MgSiO$_3$, the red points in the direction of increasing Fe content, and each curve represents a 10\% increase in Fe content up to 100\%Fe. Similarly, the blue arrow points in the direction of increasing water content at 10\% increments per curve. Solar System planets are added for comparison and plotted as purple stars.
}
\label{fig:24}
\end{figure*}

The results of the $I$ calculation are presented in Fig.\ref{fig:24}.
We noted that only TRAPPIST-1 shows intrasystem similarity in the interiors of four planets (with $I \sim$0.01-0.017), but several pairs of more massive planets also show similarity. We used the mesh of MRR curves that has a 10\% increase in each given chemical compound content from 100\%MgSiO$_3$ to 100\%Fe in the direction of increasing density and to 100\%H$_2$O in the opposite direction. In this case, the values of $I\ge$0.1 effectively put a pair of planets on curves that will be at least three curves apart and therefore differ in composition by $\ge$30\%. There are 15 pairs of planets with $I$<0.1, indicating they could have relatively similar interiors. There are 26 pairs that have $I\ge$0.1, and therefore probably differ substantially in the interior structure. This suggests that neighbouring planets with similar radii often have vastly different compositions, interior structures, and volatile contents.
\begin{figure}[ht!]
\resizebox{\hsize}{!}{
\includegraphics{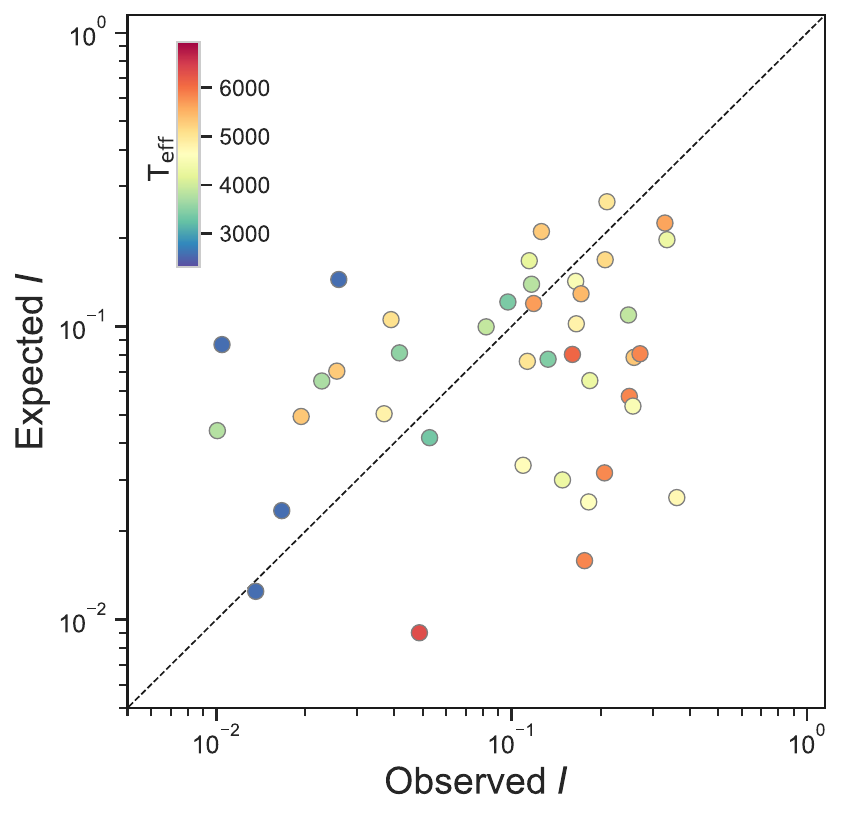}
}
\caption{Interior distance in HCR sample systems. Observed $I$ in the x-axis is plotted against expected $I$ from non-parametric MRR \citep{2018ApJ...869....5N} at the y-axis. The colours of the dots indicate stellar T$_\mathrm{eff}$ of the system's host star. The black dashed line is unity.}
\label{fig:45}
\end{figure}

We proceed with the error-accommodation routine, and this test showed that the mean value $I_{err}$ and the corresponding standard deviation for some of the pairs similar in the interior could have $I$>0.1, but for pairs with dissimilar interior $I_{err}$ can be even larger than the observed $I$. A couple of planet pairs exhibiting the greatest compositional similarity, Kepler-3 b and c and Kepler-176 c and d, have their interior distance $I_{err}$ inflated to > 0.1 due to the large observational errors.
Otherwise, the error-accommodation routine shows similar results. We tested the significance of our results by comparing them to the 10$^5$ test populations in the same way as previously and obtained $I_{sim}$. In Fig.\ref{fig:45}, we plotted observed $I$ against expected $I_{sim}$. We examined trends with host star T$_\mathrm{eff}$ and noted that cold stars show more or equal similarity in their planets' interior than expected, and hotter stars hosted planets with less similarity in the interior than expected. All results of the $I$ calculation are presented in Table \ref{table:6}, where along with the mean value of $I_{sim}$, we listed also the median values due to most distributions in the simulations being non-symmetric.

\subsection{Similarity in the systems with similar stellar properties}

\begin{figure}[ht!]
\resizebox{\hsize}{!}{\includegraphics{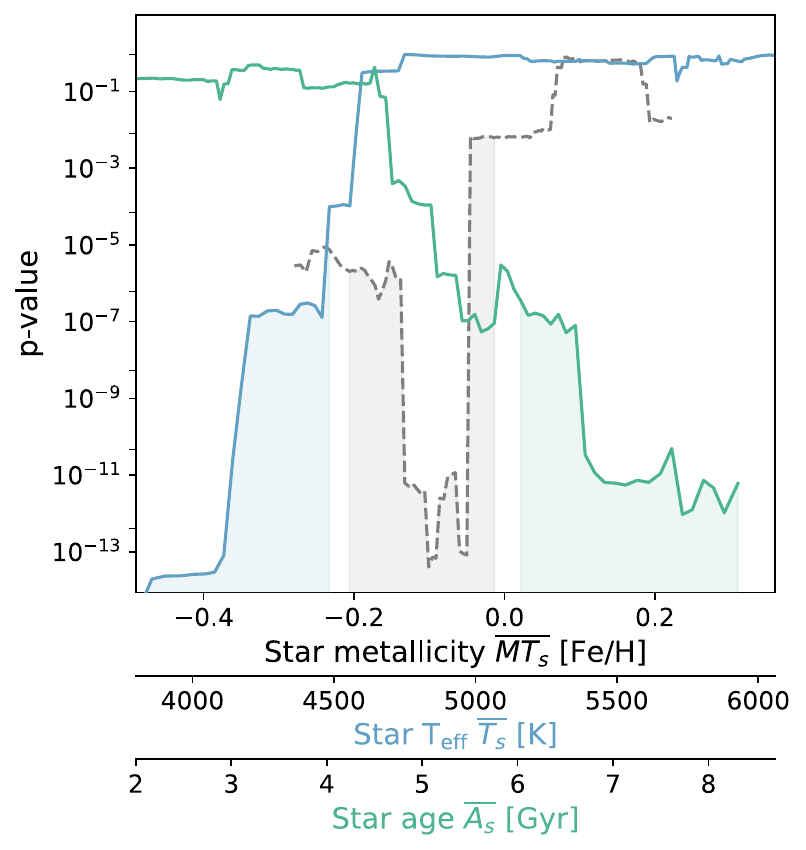}}
\caption{P-value dynamics for subsets of data points. Coloured lines indicate P-value changes during the moving window test for density depending on host star age (green) or metallicity (grey) and for masses depending on host star T$_\mathrm{eff}$ (blue). The light-coloured shades under the curves roughly indicate these subsamples in corresponding colours.}
\label{fig:241}
\end{figure}

We explored the possibility that the strength of the peas in a pod trend for adjacent planet pairs masses, radii, bulk densities and period ratios in triples depends on the properties of the host star. Such a correlation could arise from detection biases or from astrophysics. For example, \citet{zhu2020patterns} mentioned that smaller planets are more easily detected around more quiet stars. On the other hand, the host star properties, such as stellar mass or metallicity should, in principle, affect planet formation and evolution leading to certain planet interior structures as the end product. We used the moving window test and found that certain parts of the sample show signs of possible enhanced underlying correlation for masses and densities of adjacent planets. For planetary radii, the R and p-value somewhat fluctuate around the previously obtained values for the main radius subsample, giving R values close to or inside of 95\% CI, and therefore the `40 radius pairs' subsamples do not correlate any stronger than the entire dataset (see Fig.\ref{fig:23} in Appendix \ref{sec:appendix}). The same test revealed no dependency on stellar parameters for the period ratios in triples, with the maximum value of R<0.5.

Masses of planet pairs around cool host stars (T$_\mathrm{eff}$ $\in$ [2566; 4398] K) are correlated especially well (see Table \ref{table:1}), otherwise, the high R values also are typical for planet pairs around small, low mass, slightly metal-poor stars. Densities of planet pairs orbiting host stars with metallicity lower than solar (MT $\in$ [-0.210; -0.010] [Fe/H]) and of old age (A $\in$ [6.03; 12.3] Gyr) show a high correlation.
Fig.\ref{fig:241} shows p-value dynamics for these subsamples in coloured lines for density depending on host star age or metallicity and for masses depending on host star T$_\mathrm{eff}$.
Fig.\ref{fig:28b} in Appendix \ref{sec:appendix} demonstrates these subsamples along with one random run of the bootstrap from the whole subsample: for masses around cool stars, for densities around host stars with less than solar metallicity, and for densities around old stars. Additionally, we compared the resulting R values with the bootstrap test distributions in Fig.\ref{fig:29} in Appendix \ref{sec:appendix}. These R are far removed from the bootstrapped distributions and are therefore highly significant.

\subsection{Assessment of data homogeneousness}

Here we aimed at using a large and diverse sample, which is why we used the values collected by the databases from diverse sources in the literature. While this maximises the size of our sample, it also means the measurements are inhomogeneous. Homogenous stellar data could greatly help to minimise selection biases from the observations. The planetary radii uncertainties for transiting planets are dominated by the uncertainties in the stellar radii due to the transit depth $\delta = \big (R_p/R_\star \big)^2$, where $R_p$ and $R_\star$ are radii of planet and star respectively, and stellar radius determination can greatly affect the known value of the individual planet radius \citep{muirhead2012characterizing}. To investigate the impact of using a set of homogeneously determined system parameters on our statistical results, we calculated new planetary radii from
Gaia DR3 stellar data. Namely, the radii of the sample stars from FLAME and GSPPhot and associated uncertainties were retrieved.
A planet-to-star ratio for every data point of the constructed radius subsample was queried from the NASA Exoplanet Archive, and the most recently reported value was used. The majority of the Kepler planet planet-to-star radius ratios came from the DR 25 KOI catalogue \citep{hoffman2017uniform}.

From 256 initial data points for the radii correlation analysis, we recalculated radii for 227 pairs for all systems that have radius data of the star from FLAME and GSPPhot. The results for this 227 pairs sample are represented in Table \ref{table:1} under the section "Radii from Gaia DR3 correlation". 62\% of the radius pairs are above the 1:1 line, as for previous radii values, the retrieved Pearson R is lower than our previous result for the initial sample, and the p-values indicate slightly lower confidence in the results. The newly obtained radii of adjacent planets still appear to be moderately correlated, but with R$\leq$0.454 which is less than the previously obtained value. We proceed to calculate dispersion in radius $D$, using these newly obtained radii from Gaia DR3. We found that $D$ shows similarly low values for both FLAME and GSPPhot inferred radii, and the error-accommodation test shows the results very close to the obtained values. The results can be found in Table \ref{table:15} under the section "$D$ for radius from Gaia DR3". Therefore, based on these results from the Gaia homogenous sample, we consider our initial sample to be relatively homogenous in radius as well. Moreover, as we found no evidence that masses of planets in our sample determined by two different methods: RV and TTVs, are distributed differently (see Fig.\ref{fig:33} in Appendix \ref{sec:appendix}) we consider, albeit with less confidence, our sample relatively homogenous in mass.

\section{Discussion}
\label{sec:discussion}
\subsection{Radii are more uniform in pairs and systemwide than any other parameter}

We compared key properties of adjacent planets, as well as their intrasystem dispersion, in a large sample of multi-planet systems. We found that mass, density, and period ratios are generally less uniform than radius. Using the Pearson correlation test, we found moderate correlation for radii, weak correlation for densities, and no correlation for masses of adjacent planets and period ratios in triples in the entire sample. On the other hand, the dispersion metric indicates increased system-wide uniformity in mass and density compared to the null hypothesis (significant to $\sim$3$\sigma$). Moreover, among systems with high levels of radius correlation, masses and densities are not similar to nearly the same degree. The results of this study could further hint at why the MRR diagram presents a large scatter. Below we compare our results with existing observational and theoretical studies on peas in a pod patterns from the literature.

\subsubsection{Radius trend}
The radii of planet pairs in our main sample as measured by the Pearson coefficient have consistently shown medium to strong correlation (R = 0.516). They are fully consistent with those of \citet{weiss2018california} (R = 0.62 for 355 multi-planet systems) and \citet{millholland2021split} (R = 0.444 for 520 systems). Furthermore, when using planetary radii based on homogeneously derived stellar radii from Gaia DR3, we show that the radii correlation is slightly diminished (R$\geq$0.45), but still highly significant.

\subsubsection{Period ratio trend}

\citet{weiss2018california} found consecutive planets tend to have regular period spacings, as indicated by the Pearson coefficient (R = 0.46 in adjacent period ratios for a sample of 104 systems with more than 3 planets). This is apparently in tension with the lack of correlation in period ratios in our main sample. However, it is worth noting that the systems studied by \citet{weiss2018california} excluded planet pairs with period ratios > 4. As demonstrated in \citet{jiang2020orbital} and \citet{otegi2021similarity}, imposing such a restriction tends to inflate the significance of the correlations.

When focusing on compact systems with small planets where all period ratios pr $\leq$ 4 and radii R$_p\leq$ 6 R$_\oplus,$ \citet{jiang2020orbital} reported significant spacing correlation. We repeat our analysis on subsamples formed by enforcing the same upper bounds in period ratio and radii. Table \ref{table:2} presents the corresponding R and p-values, which show no evidence of significant correlation. Such discrepancy could be in part due to differences in the underlying samples: our sample is constructed from systems containing planets where at least two of them have known radii and masses, whereas other authors have generally not demanded that masses are known. On the other hand, we do observe a reduction in the system-wide dispersions for the subsamples with $pr\leq$ 4. The TTV subsample also presents lower $D$ in pr, which could be explained by the systems being preferentially near mean motion resonances \citep{jiang2020orbital}.

In planet synthesis models of \citet{mishra2021new}, the underlying population of simulated planets with pr < 4 exhibits a high degree of spacing correlation (R = 0.55). However, Kepler-like transit detection biases substantially reduce the apparent correlation to R = 0.25. Therefore, it is possible that the true spacing regularities are greater than they seem in the observed sample.

\subsubsection{Mass trend}
\label{sec:discussionmass}

Since precise masses are difficult to measure, the peas in a pod trend in mass has been more challenging to study. The absence of effective ways to deal with measurement errors in most similarity metrics means samples are often formed with strong precision constraints. This leads to limited sample sizes, and potentially biases samples towards more massive planets at a given radius. Furthermore, the use of TTV versus RV masses is a topic of ongoing debate. The RV signal induced by small planets is often less than 1 ms$^{-1}$, which is below the detection threshold of current state-of-the-art instruments and below typical noise floors from stellar activity ($\sim$2 ms$^{-1}$) and photon noise ($\sim$2 ms$^{-1}$, \citet{weiss2014mass}). The TTV method is also problematic in that it is only useful for planet pairs in near-resonance, and can overestimate masses for planets with high orbital eccentricities \citep{2014ApJ...787...80H,2021ApJ...908..114Y}.

Many authors have explored intrasystem trends in planetary mass and found evidence of relatively significant similarity (from $\sim$2.5 to 8$\sigma$). \citet{millholland2017kepler} analysed 37 systems with planet masses derived from TTV (all below 50 M$_\oplus$) and found a smaller dispersion within systems relative to random shuffling to high significance ($\sim$8$\sigma$). Citing concerns about selection biases towards resonant orbital configurations in TTV systems, \citet{Wang_2017} used 29 systems with RV planets and showed that a mass threshold (at 30 M$_\oplus$) is required to reproduce the mass uniformity to 3$\sigma$. When systems with gas giants $\gtrsim$ 100 M$_\oplus$ are included (representing half of the sample), the system-wide mass dispersion is indistinguishable from random. \citet{2022ApJ...933..162G} studied 17 systems with masses derived exclusively from RV measurements without imposing mass limits. They used the Gini index \footnote{The Gini index is known to be biased and carry large uncertainties for small data sets \citep{deltas2003small}.} as their similarity metric and found a hint of intrasystem mass uniformity, albeit only to modest significance ($\sim$2.5$\sigma$). Recently, \citet{2023MNRAS.525L..66L} computed mass dispersion for 100 systems harbouring only planets with M$_p$ < 30 M$_\oplus$ and found uniformity significant to $\sim$6$\sigma$.

For this study, we assembled a relatively large mass sample using a combination of TTV and RV measurements and lenient error precision criteria. In computing the similarity metrics, the errors are accounted for using the error-accommodation routine. Our main sample does not appear to present mass uniformity to the same extent as previous works. However, removing the systems harbouring massive planets (> 100 M$_\oplus$) from our analysis does strengthen correlations in mass. We tested this by selecting 73 pairs of planets with M$_p$ < 100 M$_\oplus$ and R$_p$ < 10 R$_\oplus$ and error precision thresholds as used for the main sample. In this subsample, the Pearson coefficients show a medium to strong correlation in radius (R = 0.69) and a weak to medium correlation in mass (R = 0.35). Correspondingly, the system-wide dispersion in mass also contracted compared to the main sample (see Table \ref{table:15}). Nevertheless, they remain far below the degree of similarity observed in radius. Intriguingly, the density correlation is weakened (R = 0.16) in the mass-restricted sample compared to the main sample, and the dispersion is not notably changed.

On the other hand, the physical motivation for excluding a certain type of planet from characterising general trends in planetary system architecture is less clear. Gas giants may play a fundamental role in sculpting planetary systems (e.g., \citet{2023A&A...674A.178B}), and large planets are commonplace, especially around FGK stars (e.g., \citet{2011arXiv1109.2497M,2019AJ....158..109H,2021ApJS..255...14F}) and including in our own system. In \citet{Wang_2017}'s entire sample, half the systems contained planets $\gtrsim$100 M$_\oplus$. Therefore, the results from our main sample are also instructive. Moreover, the population synthesis simulations of \citet{mishra2021new} did not require a mass cut-off to produce a mass correlation in adjacent planet pairs comparable to that in radius (both with R $\sim$0.6). By comparing this prediction with the observed results from our main sample, we can test the planet formation models. The fact that they contradict our observed results seems to undermine \citet{mishra2021new}'s conclusion that radii similarity is rooted in mass similarity. Nevertheless, it is also possible that the observed and theoretical samples are not directly comparable because detection biases associated with planetary mass measurements were not modelled in \citet{mishra2021new}.

\subsubsection{Taken together: Comparison with  Otegi et al. (2022)}

The most comparable study to ours is \citet{otegi2021similarity}. In a sample of 144 planet pairs from 48 multi-planet systems with well-determined masses and radii, \citet{otegi2021similarity} investigated similarities in radii, masses, as well as bulk densities, finding significant pair-wise correlations in all three parameters. However, the correlation is much weaker in mass than it is in radii and density, which is in qualitative agreement with our conclusions.

A key difference in our samples could contribute to the discrepancy in the mass correlations. Our sample is constructed using newer data and a more relaxed selection criteria in parameter errors (which we marginalise over in our analysis) and is therefore also larger (>180 systems). To test the effect of using data with the same precision as \citet{otegi2021similarity}, we repeated our analysis on a subsample selected with \citet{otegi2021similarity}'s error criteria. This resulted in a greatly strengthened density correlation, but no meaningful change to that in mass (see Table \ref{table:1} under "Error precision tests")\footnote{Indeed, \citet{otegi2021similarity} also showed that mass correlation is actually weaker among planets with more precise mass measurements.}. We suggest that recent updates to the exoplanet inventory or differences in the definitions of the similarity metrics could be responsible for the remaining discrepancy.

\citet{otegi2021similarity} found that most systems display higher dispersion in mass compared with radius, which they posited to be a consequence of densities tending to be similar in a given planetary system. We directly investigated density dispersions in all the systems (see Fig.\ref{fig:41}). We found that systems with low dispersions in mass and radius tend to have low dispersions in density, but systems with higher dispersions in mass and/or radius show a variety of density dispersions. Moreover, our findings are compatible with \citet{otegi2021similarity}'s realisation that systems with the greatest intrasystem uniformity in radius are often not the most uniform in mass, and vice versa.

Similar to \citet{Wang_2017}, \citet{otegi2021similarity} found a transition in the peas in a pod behaviour at around Saturn-mass ($\sim$100M$_\oplus$, $\sim$10R$_\oplus$), such that excluding the giant planets improved uniformity in radius and mass. This is also observed in our sample, as discussed in Section \ref{sec:discussionmass}. However, we find no evidence for an analogous behaviour in density, and it is not mentioned by \citet{otegi2021similarity}.

\subsubsection{Dependence on stellar parameters}

\citet{millholland2017kepler} stated that previous works have shown that intrasystem radius uniformity depends only weakly on stellar properties. On the contrary, \citet{otegi2021similarity} found that planetary systems around less massive stars tend to be more uniform in mass, radius, and density, since the planets tend to be less massive. In our study, we found that the correlation in radius fluctuates around R$\sim$0.5, but some of the subsamples show significantly enhanced correlations in mass (R > 0.85) for planets orbiting cool stars (T$_\mathrm{eff}$ < 4400 K) and in density for slightly metal-poor (-0.2 < [Fe/H] < 0) or older (age > 6 Gyr) host stars (see Fig.\ref{fig:29}), suggesting that planet formation mechanisms around cool, old dwarfs conspire to produce more uniform planets. It would be interesting to investigate whether this hypothesis holds in a larger sample of systems with well-determined stellar parameters.

\subsection{Light and heavy `peas' in the same `pod'}
In agreement with previous studies, we find significant intrasystem uniformity in planet radii in our sample. This trend could be due to underlying astrophysical reasons, such that similarly sized planets are common outcomes from planet formation processes in a given system (see e.g. \citet{2022arXiv220310076W}). However, these trends must be interpreted in the context of selection and observational biases associated with the transit method, which is required to measure planet radii. In addition to the effect of detection bias related to the SNR floor, as pointed out by \citet{zhu2020patterns}, we must remember that our knowledge of the planet systems is inherently incomplete.  The transit technique is only sensitive to close-in planets (i.e. `inner system architecture') that transit, which means a large part of the system architecture could be amiss \citep{2021ARA&A..59..291Z}. These missing planets could distort the similarity and spacing trends. An indication of this is the fact that, for mass and density, we find the intrasystem dispersion in systems with more than 3 planets to be greater than that in systems with more than 2 planets.

In contrast with radii, the evidence of system-wide correlation in planet masses is far more tenuous. As discussed in Section \ref{sec:discussionmass}, there have been few studies on peas in a pod trends in planet mass. While they have found correlations, the results are based on small and possibly biased samples. This is because mass measurements tend to be more difficult than radii, with larger typical errors and more susceptibility to bias.  In this work, we used a large sample of the latest available planet mass measurements in multi-planet systems, from both RV and TTV. We find that intrasystem correlation in mass is less prominent than in radius, otherwise proven uniformity in both parameters will be an important constraint for planet formation theories that model planet masses.

One reason for the difference in correlation patterns between planet radii and mass could be purely mathematical. As we imagine the planets as spherical, for a given bulk density, the radius variations are less than those in mass, and radii fluctuate in a much smaller range than masses. For example, in this study's sample, the radii of planets occupy the range of 0.28 to 13.2 R$_\oplus$ and masses those of 9.85$\times$10$^{-2}$ to 2.16$\times$10$^{3}$  M$_\oplus$, which leads to great variety in bulk densities. So one can deduce that radii will fluctuate in a much smaller range than masses in the population.
As explained in Section \ref{sec:methods}, extreme outliers can affect the behaviour of the Pearson R. Since masses scale as radius cubed, correlations in mass are more prone to being skewed by a few points on the extremity. For example, if we added the massive (and relatively highly correlated) direct imaging planet systems HR 8799, PDS 70, TYC 8988-760-1 to the sample, the Pearson coefficient for masses grows to R = 0.91, whereas it is only R = 0.65 for radius.
On the other hand, the intrasystem dispersion and parameter/interior gap calculations, two other methods we used in this work, are not affected by the range of variables due to normalisation. Using all these methods combined, we showed that parameters of the planets in the exoplanetary systems other than radius vary greatly.

A lack of mass correlation despite radius correlation can also imply non-uniformity in bulk density.  Indeed, we find only weak evidence for intrasystem correlation in bulk density. Our sample analyses showed that at a certain radius, the planet could have a variety of masses due to widely varying bulk densities.
Only a small subsample of planet pairs is highly correlated in both mass and radius (the number of pairs N$_P$=49, the number of systems N$_s$=47, the number of planets N$_p$=97) and even fewer are also similar in density (N$_P$=35, N$_s$=34, N$_p$=70). We acquired these two sub-samples by cross-checking planets belonging to HCR and HCM subsamples in the former, and all three HCR, HCM and HCD subsamples in the latter case.

Within a given system, where formation conditions are expected to be similar (especially in the inner regions probed by the compact multi-planet systems studies), one might expect planets with similar radii to also be similar in composition and interior structure, like Venus and Earth. Therefore, the fact that the exoplanets in the HCR sample tend to have dissimilar bulk densities is surprising. While bulk densities are a proxy for interior structure, their correspondence is not 1:1. To understand the degree of fundamental physical similarities between adjacent planet pairs with superficially similar radii, we locate the HCR sample on planetary interior models of \citet{zeng2016mass} in bulk density-mass space (Fig.\ref{fig:4}). Planet pairs that are alike in composition and structure should cluster along one specific interior structure contour like "strung beads". This should be especially prominent for the denser rocky planets, which do not have significant atmospheres to modify their radii. We find that, with few exceptions, the planet pairs straddle across composition contours. Indeed, apparent similarity in planet size does not mean similarity in composition or interior structure.

In their work, \citet{zeng2019growth} hypothesised that terrestrial planets, including those that are less dense than the Earth, can be divided by interior composition for two types: predominantly MgSiO$_3$ with Fe cores (rocky worlds) or predominantly H$_2$O with MgSiO$_3$ layers (water-worlds). We include these prescribed interior curves in our analyses (see Fig.\ref{fig:24}). We noted, that two adjacent planets can differ in density and mass in such a way, that one planet will belong to rocky worlds and its neighbour will belong to the water worlds. These results contradict \citet{millholland2017kepler}, as they found that astrophysical scatter in general the MRR is dominated by or at least largely linked to different systems with different available planet-forming constituents or, in other words, inter-system differences, rather than intrasystem differences in interiors of planets. On the other hand, there are several notable examples of planets exhibiting a high degree of intrasystem compositional uniformity. One is the TRAPPIST-1 planets, four of which (b, c, f, g) are in our HCR sample. Our results are consistent with \citet{agol2021refining}, who studied the TRAPPIST-1 system in detail and found that all the planets are consistent with one interior model contour: either one with depleted iron relative to Earth or an Earth-like composition with an enhancement in light elements, such as a surface water layer.

We introduced interior distance as a metric to quantify the compositional difference between similar-sized planets in the same system and suggested $I$ = 0.1 as a threshold to distinguish between planet pairs with similar and dissimilar interior structure. This interpretation compares well with \citet{2023AJ....166..137R}, who examined core/water mass fractions of planets with M$_p$ < 10 M$_\oplus$ in multi-planet systems around M dwarfs. For example, for the HD 260655 b-c pair, our calculated $I$ = 0.117 would indicate dissimilar structure. This is consistent with \citet{2023AJ....166..137R}, who determined that the system consists of a rocky and water-rich world. On the other hand, the LTT 1445A b-c pair has I = 0.053 to indicate similar structures, which is consistent with the assessment of \citet{2023AJ....166..137R} (rocky planets with comparable core-mass fractions).

The volatile-rich population in a range of Neptune and Uranus-like planets in means of radius and mass is widely represented in our HCR subsample, but their constituents are currently only theorised \citep{dorn2015can,2017A&A...597A..37D,2021ApJ...922L...4D} due to our ability to constrain interior structure is directly limited by data uncertainties and lack of such short-period planets in the Solar System. Further constructing a mesh of potential MRR based on interior structures not only for terrestrial planets but for the volatile-rich population by considering their irradiation environment and possible gaseous envelopes, following the work of \citet{zeng2019growth}, could aid in exploring the interior composition and mass range of a potential atmosphere.

\subsection{The extent of our knowledge about exoplanets}
Unlike in the Solar System where largely no radius correlation between pairs exists and high intrasystem dispersion in this parameter is present, the radius correlation in exoplanetary systems and intrasystem dispersion persists as an observational pattern. Although many authors report similar findings, the true nature of this pattern was questioned before by \citet{zhu2020patterns} who attributed it to Kepler observational biases, but also by \citet{murchikova2020peas}, who used the same sample as \citet{weiss2018california} and showed that similar patterns could arise in a mock universe with unnatural assumptions about the MRR of the planets. Since the radius is the only planetary parameter that shows a consistent correlation pattern in our entire sample and significantly lower intrasystem dispersion compared to other fundamental parameters, it is reasonable to assume that our limited knowledge of the exoplanet occurrence rates, impacted by geometric and detection biases, affects this. Currently, we can determine the radii of transiting planets only, which means we are only sensitive to planets above a certain size threshold, on relatively short orbits and are coplanar with the others in a given system. Further, the Kepler mission had its observational limitations, which we discussed previously. It has been shown that small, Earth-sized planets are very common and greatly outnumber larger planets (e.g.,\citet{petigura2013prevalence,burke2015terrestrial,fulton2017california}).

Using Kepler data and simulations, \citet{mulders2018exoplanet} predicted that at least half of known exoplanets within the habitable zone (HZ) are accompanied by non-transiting planets at shorter orbital periods. Inversely, it means that at least some of the known systems with close-in planets also are harbouring smaller planets in the HZ  and at further separation, transiting or not.
The presence of small planets will break the similarity pattern, as was discussed by \citet{zhu2020patterns}. \citet{2018AJ....156...92Z} stated that outside the Kepler period limit ($\sim$1 yr), cold giant planets preferentially coexist with inner small planets. Recently, \citet{millholland2022edge} showed that if the compact multi-planet systems with the prominent peas in a pod patterns in radius and period ratios would have additional transiting planets, consistent with these patterns and located beyond the known planets, the existing observation facilities would already be able to detect them. Thus, if these known compact systems are in fact harbouring additional planets at larger separations, these planets would need to have radii smaller by a factor of $\lesssim$ 0.5 that of super-Earth, which would be Earth-size planets. In recent work, \citet{2023A&A...676A.131H} show that typically simulated systems, formed for example around K-star, would harbour several Earth-size planets at larger separations from the star in addition to several close-in Super-Earths. The incompleteness of observed systems could be connected to our findings that systems with three or more planets show large dispersion in mass and density. Therefore, the incompleteness poses a question mark on the apparent similarity in parameters within multi-planet systems.

Finally, the measured parameters of exoplanets depend highly on the precise determination of stellar masses and radii. Gaia DR3 release provides homogeneous data for hundreds of millions of stars but has its shortcomings in characterisation of dim sources \citep{smart2021gaia,2021A&A...649A...3R}, i.e cool M- and K-dwarfs, the stars most interesting in means of finding terrestrial planets. The new Euclid ESA mission \citep{2022A&A...662A..92E}, launched in 2023, will produce a star catalogue based on Gaia data, providing more accurate stellar parameters.
Due to the limitation of the methods of exoplanet mass and radius determination, our knowledge of the entire population of exoplanets is incomplete, and it is possible that the apparent trends that we observe in known systems appear only in parts of the intrinsic distributions of fundamental parameters. Future missions such as PLATO \citep{2014ExA....38..249R} along with ongoing missions and growing knowledge about exoplanets’ properties will contribute to overcoming these obstacles.

\section{Conclusions}
\label{sec:conclusions}

In this paper, we study the `peas in a pod' pattern of radii, masses, densities of adjacent planet pairs, and orbital period ratios of triples belonging to the same system. Our aim is to investigate whether or not the apparent similarity in radius implies similarity in mass, density, and interior structure. We were motivated to revisit these patterns by the fact that the masses of many planets have recently been determined. As in previous works, we find that planet radii within the same system consistently and robustly show similarity as opposed to a random arrangement. We additionally tested radius similarity using the homogenous dataset of stellar radii from Gaia DR3 and come to the same conclusion. This is probably due to one or more of the following reasons: (1) the physical processes creating the planets in the same system with similar radii are probably the same; (2) there are physical constraints that mean that radius will vary less than other planet parameters, (e.g. uniformity of the protoplanetary disc could lead to similar density and interior structure of produced planets); or (3) the current sample is still not reflecting the entire exoplanets population.

Planetary masses within a system are not as uniform as radii and reflect the overall scatter in the MRR for the whole variety of exoplanets, covering a wide range of bulk densities. Using a fine grid of theoretical MRR compositional curves, we investigated the extent to which the peas in a pod pattern propagates into the chemical compositions of planets in compact multiple systems and find that apparently similar-sized planets are often not alike in composition or interior structure. Two adjacent planets similar in radius can belong to the rocky- and water-world categories, respectively. In other words, even within the same system, it cannot be assumed that similarity in radius necessarily implies similarity in mass, density, composition, or internal structure. The principle of separating the data by prescribed interior structures is preferable to separating by the density of a uniform sphere because the constant density would not account for the self-compression of an object. As we gather more information on the detailed nature of planets in these systems, characterisation of this population beyond the superficial parameters of size and mass will help to shed light on the physical origin of these intriguing intrasystem trends, or the lack thereof.

Interestingly, the degree of likeness in adjacent planet masses seems to depend on the stellar properties of the host. For example, small, less metal-rich stars tend to host planets with similar mass and density, which probably suggests there is comparatively less building material around such stars with which to form large planets, meaning that all the formed and detected planets tend to be small and more similarly sized.
The similarities in masses and densities within one system could originate from the epoch of planet formation, potentially prescribing a certain amount of mass, and the constituents of a protoplanetary disc, as well as its structure, shape the architecture of the system. Further study of the similarities in mass and density among systems with similar of a host-star parameters  using a much larger sample could further our understanding of the planets' internal structure and the processes of planet formation.

\begin{acknowledgements}
We acknowledge financial support from the Research Council of Norway (RCN), through its Centres of Excellence funding scheme, project 223272 (CEED) and project number 332523 (PHAB). We thank Pavel Dubrovine (PHAB), Anders Erikson (German Aerospace Centre, Berlin-Adlershof, Germany), Mathew Domeier (PHAB), and Alexey Pankine (Space Science Institute, USA) for the fruitful discussion. We thank the anonymous reviewer for the thoughtful feedback that helped to improve this manuscript.
\end{acknowledgements}

\bibliography{bibliography}

\begin{thebibliography}{80}
\expandafter\ifx\csname natexlab\endcsname\relax\def\natexlab#1{#1}\fi

\bibitem[{Agol {et~al.}(2021)Agol, Dorn, Grimm, Turbet, Ducrot, Delrez, Gillon,
  Demory, Burdanov, Barkaoui, {et~al.}}]{agol2021refining}
Agol, E., Dorn, C., Grimm, S.~L., {et~al.} 2021, The planetary science journal,
  2, 1

\bibitem[{Akeson {et~al.}(2013)Akeson, Chen, Ciardi, Crane, Good, Harbut,
  Jackson, Kane, Laity, Leifer, {et~al.}}]{akeson2013nasa}
Akeson, R., Chen, X., Ciardi, D., {et~al.} 2013, Publications of the
  Astronomical Society of the Pacific, 125, 989

\bibitem[{{Andrae} {et~al.}(2023){Andrae}, {Fouesneau}, {Sordo},
  {Bailer-Jones}, {Dharmawardena}, {Rybizki}, {De Angeli}, {Lindstr{\o}m},
  {Marshall}, {Drimmel}, {Korn}, {Soubiran}, {Brouillet}, {Casamiquela}, {Rix},
  {Abreu Aramburu}, {{\'A}lvarez}, {Bakker}, {Bellas-Velidis}, {Bijaoui},
  {Brugaletta}, {Burlacu}, {Carballo}, {Chaoul}, {Chiavassa}, {Contursi},
  {Cooper}, {Creevey}, {Dafonte}, {Dapergolas}, {de Laverny}, {Delchambre},
  {Demouchy}, {Edvardsson}, {Fr{\'e}mat}, {Garabato}, {Garc{\'\i}a-Lario},
  {Garc{\'\i}a-Torres}, {Gavel}, {Gomez}, {Gonz{\'a}lez-Santamar{\'\i}a},
  {Hatzidimitriou}, {Heiter}, {Jean-Antoine Piccolo}, {Kontizas}, {Kordopatis},
  {Lanzafame}, {Lebreton}, {Licata}, {Livanou}, {Lobel}, {Lorca}, {Magdaleno
  Romeo}, {Manteiga}, {Marocco}, {Mary}, {Nicolas}, {Ordenovic}, {Pailler},
  {Palicio}, {Pallas-Quintela}, {Panem}, {Pichon}, {Poggio}, {Recio-Blanco},
  {Riclet}, {Robin}, {Santove{\~n}a}, {Sarro}, {Schultheis}, {Segol},
  {Silvelo}, {Slezak}, {Smart}, {S{\"u}veges}, {Th{\'e}venin}, {Torralba
  Elipe}, {Ulla}, {Utrilla}, {Vallenari}, {van Dillen}, {Zhao}, \&
  {Zorec}}]{andrae2022gaia}
{Andrae}, R., {Fouesneau}, M., {Sordo}, R., {et~al.} 2023, \aap, 674, A27

\bibitem[{Armstrong {et~al.}(2020)Armstrong, Lopez, Adibekyan, Booth, Bryant,
  Collins, Deleuil, Emsenhuber, Huang, King, {et~al.}}]{armstrong2020remnant}
Armstrong, D.~J., Lopez, T.~A., Adibekyan, V., {et~al.} 2020, Nature, 583, 39

\bibitem[{Bashi {et~al.}(2017)Bashi, Helled, Zucker, \&
  Mordasini}]{bashi2017two}
Bashi, D., Helled, R., Zucker, S., \& Mordasini, C. 2017, Astronomy \&
  Astrophysics, 604, A83

\bibitem[{Bishara \& Hittner(2015)}]{bishara2015reducing}
Bishara, A.~J. \& Hittner, J.~B. 2015, Educational and psychological
  measurement, 75, 785

\bibitem[{{Bitsch} \& {Izidoro}(2023)}]{2023A&A...674A.178B}
{Bitsch}, B. \& {Izidoro}, A. 2023, \aap, 674, A178

\bibitem[{Borucki {et~al.}(2010)Borucki, Koch, Basri, Batalha, Brown, Caldwell,
  Caldwell, Christensen-Dalsgaard, Cochran, DeVore,
  {et~al.}}]{borucki2010kepler}
Borucki, W.~J., Koch, D., Basri, G., {et~al.} 2010, Science, 327, 977

\bibitem[{Burke {et~al.}(2015)Burke, Christiansen, Mullally, Seader, Huber,
  Rowe, Coughlin, Thompson, Catanzarite, Clarke,
  {et~al.}}]{burke2015terrestrial}
Burke, C.~J., Christiansen, J.~L., Mullally, F., {et~al.} 2015, The
  Astrophysical Journal, 809, 8

\bibitem[{{Carter} {et~al.}(2012){Carter}, {Agol}, {Chaplin}, {Basu},
  {Bedding}, {Buchhave}, {Christensen-Dalsgaard}, {Deck}, {Elsworth},
  {Fabrycky}, {Ford}, {Fortney}, {Hale}, {Handberg}, {Hekker}, {Holman},
  {Huber}, {Karoff}, {Kawaler}, {Kjeldsen}, {Lissauer}, {Lopez}, {Lund},
  {Lundkvist}, {Metcalfe}, {Miglio}, {Rogers}, {Stello}, {Borucki}, {Bryson},
  {Christiansen}, {Cochran}, {Geary}, {Gilliland}, {Haas}, {Hall}, {Howard},
  {Jenkins}, {Klaus}, {Koch}, {Latham}, {MacQueen}, {Sasselov}, {Steffen},
  {Twicken}, \& {Winn}}]{2012Sci...337..556C}
{Carter}, J.~A., {Agol}, E., {Chaplin}, W.~J., {et~al.} 2012, Science, 337, 556

\bibitem[{Chen \& Kipping(2016)}]{chen2016probabilistic}
Chen, J. \& Kipping, D. 2016, The Astrophysical Journal, 834, 17

\bibitem[{{Creevey} {et~al.}(2023){Creevey}, {Sordo}, {Pailler}, {Fr{\'e}mat},
  {Heiter}, {Th{\'e}venin}, {Andrae}, {Fouesneau}, {Lobel}, {Bailer-Jones},
  {Garabato}, {Bellas-Velidis}, {Brugaletta}, {Lorca}, {Ordenovic}, {Palicio},
  {Sarro}, {Delchambre}, {Drimmel}, {Rybizki}, {Torralba Elipe}, {Korn},
  {Recio-Blanco}, {Schultheis}, {De Angeli}, {Montegriffo}, {Abreu Aramburu},
  {Accart}, {{\'A}lvarez}, {Bakker}, {Brouillet}, {Burlacu}, {Carballo},
  {Casamiquela}, {Chiavassa}, {Contursi}, {Cooper}, {Dafonte}, {Dapergolas},
  {de Laverny}, {Dharmawardena}, {Edvardsson}, {Le Fustec},
  {Garc{\'\i}a-Lario}, {Garc{\'\i}a-Torres}, {Gomez},
  {Gonz{\'a}lez-Santamar{\'\i}a}, {Hatzidimitriou}, {Jean-Antoine Piccolo},
  {Kontiza}, {Kordopatis}, {Lanzafame}, {Lebreton}, {Licata}, {Lindstr{\o}m},
  {Livanou}, {Magdaleno Romeo}, {Manteiga}, {Marocco}, {Marshall}, {Mary},
  {Nicolas}, {Pallas-Quintela}, {Panem}, {Pichon}, {Poggio}, {Riclet}, {Robin},
  {Santove{\~n}a}, {Silvelo}, {Slezak}, {Smart}, {Soubiran}, {S{\"u}veges},
  {Ulla}, {Utrilla}, {Vallenari}, {Zhao}, {Zorec}, {Barrado}, {Bijaoui},
  {Bouret}, {Blomme}, {Brott}, {Cassisi}, {Kochukhov}, {Martayan}, {Shulyak},
  \& {Silvester}}]{2022arXiv220605864C}
{Creevey}, O.~L., {Sordo}, R., {Pailler}, F., {et~al.} 2023, \aap, 674, A26

\bibitem[{Deck {et~al.}(2013)Deck, Payne, \& Holman}]{deck2013first}
Deck, K.~M., Payne, M., \& Holman, M.~J. 2013, The Astrophysical Journal, 774,
  129

\bibitem[{Deltas(2003)}]{deltas2003small}
Deltas, G. 2003, Review of economics and statistics, 85, 226

\bibitem[{Demory {et~al.}(2016)Demory, Gillon, De~Wit, Madhusudhan, Bolmont,
  Heng, Kataria, Lewis, Hu, Krick, {et~al.}}]{demory2016map}
Demory, B.-O., Gillon, M., De~Wit, J., {et~al.} 2016, Nature, 532, 207

\bibitem[{Dorn {et~al.}(2015)Dorn, Khan, Heng, Connolly, Alibert, Benz, \&
  Tackley}]{dorn2015can}
Dorn, C., Khan, A., Heng, K., {et~al.} 2015, Astronomy \& Astrophysics, 577,
  A83

\bibitem[{{Dorn} \& {Lichtenberg}(2021)}]{2021ApJ...922L...4D}
{Dorn}, C. \& {Lichtenberg}, T. 2021, \apjl, 922, L4

\bibitem[{{Dorn} {et~al.}(2017){Dorn}, {Venturini}, {Khan}, {Heng}, {Alibert},
  {Helled}, {Rivoldini}, \& {Benz}}]{2017A&A...597A..37D}
{Dorn}, C., {Venturini}, J., {Khan}, A., {et~al.} 2017, \aap, 597, A37

\bibitem[{Espinoza {et~al.}(2020)Espinoza, Brahm, Henning, Jord{\'a}n, Dorn,
  Rojas, Sarkis, Kossakowski, Schlecker, D{\'\i}az, {et~al.}}]{espinoza2020hd}
Espinoza, N., Brahm, R., Henning, T., {et~al.} 2020, Monthly Notices of the
  Royal Astronomical Society, 491, 2982

\bibitem[{{Euclid Collaboration} {et~al.}(2022){Euclid Collaboration},
  {Schirmer}, {Jahnke}, {Seidel}, {Aussel}, {Bodendorf}, {Grupp}, {Hormuth},
  {Wachter}, {Appleton}, {Barbier}, {Brinchmann}, {Carrasco}, {Castander},
  {Coupon}, {De Paolis}, {Franco}, {Ganga}, {Hudelot}, {Jullo}, {Lan{\c{c}}on},
  {Nucita}, {Paltani}, {Smadja}, {Strafella}, {Venancio}, {Weiler}, {Amara},
  {Auphan}, {Auricchio}, {Balestra}, {Bender}, {Bonino}, {Branchini},
  {Brescia}, {Capobianco}, {Carbone}, {Carretero}, {Casas}, {Castellano},
  {Cavuoti}, {Cimatti}, {Cledassou}, {Congedo}, {Conselice}, {Conversi},
  {Copin}, {Corcione}, {Costille}, {Courbin}, {Da Silva}, {Degaudenzi},
  {Douspis}, {Dubath}, {Dupac}, {Dusini}, {Ealet}, {Farrens}, {Ferriol},
  {Fosalba}, {Frailis}, {Franceschi}, {Franzetti}, {Fumana}, {Garilli},
  {Gillard}, {Gillis}, {Giocoli}, {Grazian}, {Guzzo}, {Haugan}, {Hoekstra},
  {Holmes}, {Hornstrup}, {K{\"u}mmel}, {Kermiche}, {Kiessling}, {Kilbinger},
  {Kitching}, {Kohley}, {Kunz}, {Kurki-Suonio}, {Laureijs}, {Ligori}, {Lilje},
  {Lloro}, {Maciaszek}, {Maiorano}, {Mansutti}, {Marggraf}, {Markovic},
  {Marulli}, {Massey}, {Maurogordato}, {Mellier}, {Meneghetti}, {Merlin},
  {Meylan}, {Moresco}, {Moscardini}, {Munari}, {Nakajima}, {Nichol}, {Niemi},
  {Padilla}, {Pasian}, {Pedersen}, {Percival}, {Pettorino}, {Pires}, {Poncet},
  {Popa}, {Pozzetti}, {Prieto}, {Raison}, {Rhodes}, {Rix}, {Roncarelli},
  {Rossetti}, {Saglia}, {Sartoris}, {Scaramella}, {Schneider}, {Secroun},
  {Serrano}, {Sirignano}, {Sirri}, {Stanco}, {Tallada-Cresp{\'\i}}, {Taylor},
  {Teplitz}, {Tereno}, {Toledo-Moreo}, {Torradeflot}, {Trifoglio}, {Valentijn},
  {Valenziano}, {Wang}, {Weller}, {Zamorani}, {Zoubian}, {Andreon}, {Bardelli},
  {Boucaud}, {Camera}, {Farinelli}, {Graci{\'a}-Carpio}, {Maino}, {Medinaceli},
  {Mei}, {Morisset}, {Polenta}, {Renzi}, {Romelli}, {Tenti}, {Vassallo},
  {Zacchei}, {Zucca}, {Baccigalupi}, {Balaguera-Antol{\'\i}nez}, {Biviano},
  {Blanchard}, {Borgani}, {Bozzo}, {Burigana}, {Cabanac}, {Cappi}, {Carvalho},
  {Casas}, {Castignani}, {Colodro-Conde}, {Cooray}, {Courtois}, {Crocce},
  {Cuby}, {Davini}, {de la Torre}, {Di Ferdinando}, {Escartin}, {Farina},
  {Ferreira}, {Finelli}, {Fotopoulou}, {Galeotta}, {Garcia-Bellido},
  {Gaztanaga}, {George}, {Gozaliasl}, {Hook}, {Ili{\'c}}, {Kansal},
  {Kashlinsky}, {Keihanen}, {Kirkpatrick}, {Lindholm}, {Mainetti}, {Maoli},
  {Martinelli}, {Martinet}, {Maturi}, {Mauri}, {McCracken}, {Metcalf},
  {Monaco}, {Morgante}, {Nightingale}, {Patrizii}, {Peel}, {Popa}, {Porciani},
  {Potter}, {Reimberg}, {Riccio}, {S{\'a}nchez}, {Sapone}, {Scottez},
  {Sefusatti}, {Teyssier}, {Tutusaus}, {Valieri}, {Valiviita}, {Viel}, \&
  {Hildebrandt}}]{2022A&A...662A..92E}
{Euclid Collaboration}, {Schirmer}, M., {Jahnke}, K., {et~al.} 2022, \aap, 662,
  A92

\bibitem[{Evans {et~al.}(2016)Evans, Sing, Wakeford, Nikolov, Ballester,
  Drummond, Kataria, Gibson, Amundsen, \& Spake}]{evans2016detection}
Evans, T.~M., Sing, D.~K., Wakeford, H.~R., {et~al.} 2016, The Astrophysical
  Journal Letters, 822, L4

\bibitem[{Fabrycky {et~al.}(2014)Fabrycky, Lissauer, Ragozzine, Rowe, Steffen,
  Agol, Barclay, Batalha, Borucki, Ciardi, {et~al.}}]{fabrycky2014architecture}
Fabrycky, D.~C., Lissauer, J.~J., Ragozzine, D., {et~al.} 2014, The
  Astrophysical Journal, 790, 146

\bibitem[{Fisher(1915)}]{10.2307/2331838}
Fisher, R.~A. 1915, Biometrika, 10, 507

\bibitem[{{Fouesneau} {et~al.}(2023){Fouesneau}, {Fr{\'e}mat}, {Andrae},
  {Korn}, {Soubiran}, {Kordopatis}, {Vallenari}, {Heiter}, {Creevey}, {Sarro},
  {de Laverny}, {Lanzafame}, {Lobel}, {Sordo}, {Rybizki}, {Slezak},
  {{\'A}lvarez}, {Drimmel}, {Garabato}, {Delchambre}, {Bailer-Jones},
  {Hatzidimitriou}, {Lorca}, {Le Fustec}, {Pailler}, {Mary}, {Robin},
  {Utrilla}, {Abreu Aramburu}, {Bakker}, {Bellas-Velidis}, {Bijaoui}, {Blomme},
  {Bouret}, {Brouillet}, {Brugaletta}, {Burlacu}, {Carballo}, {Casamiquela},
  {Chaoul}, {Chiavassa}, {Contursi}, {Cooper}, {Dafonte}, {Demouchy},
  {Dharmawardena}, {Garc{\'\i}a-Lario}, {Garc{\'\i}a-Torres}, {Gomez},
  {Gonz{\'a}lez-Santamar{\'\i}a}, {Jean-Antoine Piccolo}, {Kontizas},
  {Lebreton}, {Licata}, {Lindstr{\o}m}, {Livanou}, {Magdaleno Romeo},
  {Manteiga}, {Marocco}, {Martayan}, {Marshall}, {Nicolas}, {Ordenovic},
  {Palicio}, {Pallas-Quintela}, {Pichon}, {Poggio}, {Recio-Blanco}, {Riclet},
  {Santove{\~n}a}, {Schultheis}, {Segol}, {Silvelo}, {Smart}, {S{\"u}veges},
  {Th{\'e}venin}, {Torralba Elipe}, {Ulla}, {van Dillen}, {Zhao}, \&
  {Zorec}}]{fouesneau2022gaia}
{Fouesneau}, M., {Fr{\'e}mat}, Y., {Andrae}, R., {et~al.} 2023, \aap, 674, A28

\bibitem[{Fulton {et~al.}(2017)Fulton, Petigura, Howard, Isaacson, Marcy,
  Cargile, Hebb, Weiss, Johnson, Morton, {et~al.}}]{fulton2017california}
Fulton, B.~J., Petigura, E.~A., Howard, A.~W., {et~al.} 2017, The Astronomical
  Journal, 154, 109

\bibitem[{{Fulton} {et~al.}(2021){Fulton}, {Rosenthal}, {Hirsch}, {Isaacson},
  {Howard}, {Dedrick}, {Sherstyuk}, {Blunt}, {Petigura}, {Knutson}, {Behmard},
  {Chontos}, {Crepp}, {Crossfield}, {Dalba}, {Fischer}, {Henry}, {Kane},
  {Kosiarek}, {Marcy}, {Rubenzahl}, {Weiss}, \& {Wright}}]{2021ApJS..255...14F}
{Fulton}, B.~J., {Rosenthal}, L.~J., {Hirsch}, L.~A., {et~al.} 2021, \apjs,
  255, 14

\bibitem[{Galicher {et~al.}(2016)Galicher, Marois, Macintosh, Zuckerman,
  Barman, Konopacky, Song, Patience, Lafreniere, Doyon,
  {et~al.}}]{galicher2016international}
Galicher, R., Marois, C., Macintosh, B., {et~al.} 2016, Astronomy \&
  Astrophysics, 594, A63

\bibitem[{{Goldberg} \& {Batygin}(2022)}]{2022AJ....163..201G}
{Goldberg}, M. \& {Batygin}, K. 2022, \aj, 163, 201

\bibitem[{{Goyal} \& {Wang}(2022)}]{2022ApJ...933..162G}
{Goyal}, A.~V. \& {Wang}, S. 2022, \apj, 933, 162

\bibitem[{{Hadden} \& {Lithwick}(2014)}]{2014ApJ...787...80H}
{Hadden}, S. \& {Lithwick}, Y. 2014, \apj, 787, 80

\bibitem[{{Hatalova} {et~al.}(2023){Hatalova}, {Brasser}, {Mamonova}, \&
  {Werner}}]{2023A&A...676A.131H}
{Hatalova}, P., {Brasser}, R., {Mamonova}, E., \& {Werner}, S.~C. 2023, \aap,
  676, A131

\bibitem[{{He} {et~al.}(2019){He}, {Ford}, \&
  {Ragozzine}}]{2019MNRAS.490.4575H}
{He}, M.~Y., {Ford}, E.~B., \& {Ragozzine}, D. 2019, \mnras, 490, 4575

\bibitem[{{He} {et~al.}(2020){He}, {Ford}, {Ragozzine}, \&
  {Carrera}}]{2020AJ....160..276H}
{He}, M.~Y., {Ford}, E.~B., {Ragozzine}, D., \& {Carrera}, D. 2020, \aj, 160,
  276

\bibitem[{Hoeijmakers {et~al.}(2018)Hoeijmakers, Ehrenreich, Heng, Kitzmann,
  Grimm, Allart, Deitrick, Wyttenbach, Oreshenko, Pino,
  {et~al.}}]{hoeijmakers2018atomic}
Hoeijmakers, H.~J., Ehrenreich, D., Heng, K., {et~al.} 2018, Nature, 560, 453

\bibitem[{Hoffman {et~al.}(2017)Hoffman, Rowe, {et~al.}}]{hoffman2017uniform}
Hoffman, K., Rowe, J.~F., {et~al.} 2017, Kepler Science Document
  KSCI-19113-001, 21

\bibitem[{Howell {et~al.}(2014)Howell, Sobeck, Haas, Still, Barclay, Mullally,
  Troeltzsch, Aigrain, Bryson, Caldwell, {et~al.}}]{howell2014k2}
Howell, S.~B., Sobeck, C., Haas, M., {et~al.} 2014, Publications of the
  Astronomical Society of the Pacific, 126, 398

\bibitem[{{Hsu} {et~al.}(2019){Hsu}, {Ford}, {Ragozzine}, \&
  {Ashby}}]{2019AJ....158..109H}
{Hsu}, D.~C., {Ford}, E.~B., {Ragozzine}, D., \& {Ashby}, K. 2019, \aj, 158,
  109

\bibitem[{Jiang {et~al.}(2020)Jiang, Xie, \& Zhou}]{jiang2020orbital}
Jiang, C.-F., Xie, J.-W., \& Zhou, J.-L. 2020, The Astronomical Journal, 160,
  180

\bibitem[{{Lammers} {et~al.}(2023){Lammers}, {Hadden}, \&
  {Murray}}]{2023MNRAS.525L..66L}
{Lammers}, C., {Hadden}, S., \& {Murray}, N. 2023, \mnras, 525, L66

\bibitem[{Lissauer {et~al.}(2011)Lissauer, Ragozzine, Fabrycky, Steffen, Ford,
  Jenkins, Shporer, Holman, Rowe, Quintana,
  {et~al.}}]{lissauer2011architecture}
Lissauer, J.~J., Ragozzine, D., Fabrycky, D.~C., {et~al.} 2011, The
  Astrophysical Journal Supplement Series, 197, 8

\bibitem[{{Luque} \& {Pall{\'e}}(2022)}]{2022Sci...377.1211L}
{Luque}, R. \& {Pall{\'e}}, E. 2022, Science, 377, 1211

\bibitem[{{Mayor} {et~al.}(2011){Mayor}, {Marmier}, {Lovis}, {Udry},
  {S{\'e}gransan}, {Pepe}, {Benz}, {Bertaux}, {Bouchy}, {Dumusque}, {Lo Curto},
  {Mordasini}, {Queloz}, \& {Santos}}]{2011arXiv1109.2497M}
{Mayor}, M., {Marmier}, M., {Lovis}, C., {et~al.} 2011, arXiv e-prints,
  arXiv:1109.2497

\bibitem[{Millholland {et~al.}(2017)Millholland, Wang, \&
  Laughlin}]{millholland2017kepler}
Millholland, S., Wang, S., \& Laughlin, G. 2017, The Astrophysical Journal
  Letters, 849, L33

\bibitem[{Millholland {et~al.}(2022)Millholland, He, \&
  Zink}]{millholland2022edge}
Millholland, S.~C., He, M.~Y., \& Zink, J.~K. 2022, The Astronomical Journal,
  164, 72

\bibitem[{Millholland \& Winn(2021)}]{millholland2021split}
Millholland, S.~C. \& Winn, J.~N. 2021, The Astrophysical Journal Letters, 920,
  L34

\bibitem[{Mishra {et~al.}(2021)Mishra, Alibert, Leleu, Emsenhuber, Mordasini,
  Burn, Udry, \& Benz}]{mishra2021new}
Mishra, L., Alibert, Y., Leleu, A., {et~al.} 2021, Astronomy and astrophysics
  (Berlin), 656, A74

\bibitem[{Muirhead {et~al.}(2012)Muirhead, Hamren, Schlawin, Rojas-Ayala,
  Covey, \& Lloyd}]{muirhead2012characterizing}
Muirhead, P.~S., Hamren, K., Schlawin, E., {et~al.} 2012, The Astrophysical
  Journal Letters, 750, L37

\bibitem[{Mulders {et~al.}(2018)Mulders, Pascucci, Apai, \&
  Ciesla}]{mulders2018exoplanet}
Mulders, G.~D., Pascucci, I., Apai, D., \& Ciesla, F.~J. 2018, The Astronomical
  Journal, 156, 24

\bibitem[{Murchikova \& Tremaine(2020)}]{murchikova2020peas}
Murchikova, L. \& Tremaine, S. 2020, The Astronomical Journal, 160, 160

\bibitem[{{Ning} {et~al.}(2018){Ning}, {Wolfgang}, \&
  {Ghosh}}]{2018ApJ...869....5N}
{Ning}, B., {Wolfgang}, A., \& {Ghosh}, S. 2018, \apj, 869, 5

\bibitem[{Otegi {et~al.}(2020)Otegi, Bouchy, \& Helled}]{otegi2020revisited}
Otegi, J., Bouchy, F., \& Helled, R. 2020, Astronomy \& Astrophysics, 634, A43

\bibitem[{Otegi {et~al.}(2022)Otegi, Helled, \& Bouchy}]{otegi2021similarity}
Otegi, J., Helled, R., \& Bouchy, F. 2022, Astronomy \& Astrophysics, 658, A107

\bibitem[{Pearson(1905)}]{pearson1905general}
Pearson, K. 1905, On the general theory of skew correlation and non-linear
  regression (Dulau and Company)

\bibitem[{Petigura {et~al.}(2013)Petigura, Howard, \&
  Marcy}]{petigura2013prevalence}
Petigura, E.~A., Howard, A.~W., \& Marcy, G.~W. 2013, Proceedings of the
  National Academy of Sciences, 110, 19273

\bibitem[{Queloz \& Mayor(1995)}]{QuelozDidier1995AJct}
Queloz, D. \& Mayor, M. 1995, Nature (London), 378, 355

\bibitem[{{Rauer} {et~al.}(2014){Rauer}, {Catala}, {Aerts}, {Appourchaux},
  {Benz}, {Brandeker}, {Christensen-Dalsgaard}, {Deleuil}, {Gizon}, {Goupil},
  {G{\"u}del}, {Janot-Pacheco}, {Mas-Hesse}, {Pagano}, {Piotto}, {Pollacco},
  {Santos}, {Smith}, {Su{\'a}rez}, {Szab{\'o}}, {Udry}, {Adibekyan}, {Alibert},
  {Almenara}, {Amaro-Seoane}, {Eiff}, {Asplund}, {Antonello}, {Barnes},
  {Baudin}, {Belkacem}, {Bergemann}, {Bihain}, {Birch}, {Bonfils}, {Boisse},
  {Bonomo}, {Borsa}, {Brand{\~a}o}, {Brocato}, {Brun}, {Burleigh}, {Burston},
  {Cabrera}, {Cassisi}, {Chaplin}, {Charpinet}, {Chiappini}, {Church},
  {Csizmadia}, {Cunha}, {Damasso}, {Davies}, {Deeg}, {D{\'\i}az}, {Dreizler},
  {Dreyer}, {Eggenberger}, {Ehrenreich}, {Eigm{\"u}ller}, {Erikson}, {Farmer},
  {Feltzing}, {de Oliveira Fialho}, {Figueira}, {Forveille}, {Fridlund},
  {Garc{\'\i}a}, {Giommi}, {Giuffrida}, {Godolt}, {Gomes da Silva}, {Granzer},
  {Grenfell}, {Grotsch-Noels}, {G{\"u}nther}, {Haswell}, {Hatzes},
  {H{\'e}brard}, {Hekker}, {Helled}, {Heng}, {Jenkins}, {Johansen},
  {Khodachenko}, {Kislyakova}, {Kley}, {Kolb}, {Krivova}, {Kupka}, {Lammer},
  {Lanza}, {Lebreton}, {Magrin}, {Marcos-Arenal}, {Marrese}, {Marques},
  {Martins}, {Mathis}, {Mathur}, {Messina}, {Miglio}, {Montalban}, {Montalto},
  {Monteiro}, {Moradi}, {Moravveji}, {Mordasini}, {Morel}, {Mortier},
  {Nascimbeni}, {Nelson}, {Nielsen}, {Noack}, {Norton}, {Ofir}, {Oshagh},
  {Ouazzani}, {P{\'a}pics}, {Parro}, {Petit}, {Plez}, {Poretti}, {Quirrenbach},
  {Ragazzoni}, {Raimondo}, {Rainer}, {Reese}, {Redmer}, {Reffert},
  {Rojas-Ayala}, {Roxburgh}, {Salmon}, {Santerne}, {Schneider}, {Schou},
  {Schuh}, {Schunker}, {Silva-Valio}, {Silvotti}, {Skillen}, {Snellen}, {Sohl},
  {Sousa}, {Sozzetti}, {Stello}, {Strassmeier}, {{\v{S}}vanda}, {Szab{\'o}},
  {Tkachenko}, {Valencia}, {Van Grootel}, {Vauclair}, {Ventura}, {Wagner},
  {Walton}, {Weingrill}, {Werner}, {Wheatley}, \&
  {Zwintz}}]{2014ExA....38..249R}
{Rauer}, H., {Catala}, C., {Aerts}, C., {et~al.} 2014, Experimental Astronomy,
  38, 249

\bibitem[{{Recio-Blanco} {et~al.}(2023){Recio-Blanco}, {de Laverny}, {Palicio},
  {Kordopatis}, {{\'A}lvarez}, {Schultheis}, {Contursi}, {Zhao}, {Torralba
  Elipe}, {Ordenovic}, {Manteiga}, {Dafonte}, {Oreshina-Slezak}, {Bijaoui},
  {Fr{\'e}mat}, {Seabroke}, {Pailler}, {Spitoni}, {Poggio}, {Creevey}, {Abreu
  Aramburu}, {Accart}, {Andrae}, {Bailer-Jones}, {Bellas-Velidis}, {Brouillet},
  {Brugaletta}, {Burlacu}, {Carballo}, {Casamiquela}, {Chiavassa}, {Cooper},
  {Dapergolas}, {Delchambre}, {Dharmawardena}, {Drimmel}, {Edvardsson},
  {Fouesneau}, {Garabato}, {Garc{\'\i}a-Lario}, {Garc{\'\i}a-Torres}, {Gavel},
  {Gomez}, {Gonz{\'a}lez-Santamar{\'\i}a}, {Hatzidimitriou}, {Heiter},
  {Jean-Antoine Piccolo}, {Kontizas}, {Korn}, {Lanzafame}, {Lebreton}, {Le
  Fustec}, {Licata}, {Lindstr{\o}m}, {Livanou}, {Lobel}, {Lorca}, {Magdaleno
  Romeo}, {Marocco}, {Marshall}, {Mary}, {Nicolas}, {Pallas-Quintela}, {Panem},
  {Pichon}, {Riclet}, {Robin}, {Rybizki}, {Santove{\~n}a}, {Silvelo}, {Smart},
  {Sarro}, {Sordo}, {Soubiran}, {S{\"u}veges}, {Ulla}, {Vallenari}, {Zorec},
  {Utrilla}, \& {Bakker}}]{recio2022gaia}
{Recio-Blanco}, A., {de Laverny}, P., {Palicio}, P.~A., {et~al.} 2023, \aap,
  674, A29

\bibitem[{Ricker {et~al.}(2016)Ricker, Vanderspek, Winn, Seager,
  Berta-Thompson, Levine, Villasenor, Latham, Charbonneau, Holman,
  {et~al.}}]{ricker2016transiting}
Ricker, G.~R., Vanderspek, R., Winn, J., {et~al.} 2016, in Space Telescopes and
  Instrumentation 2016: Optical, Infrared, and Millimeter Wave, Vol. 9904,
  SPIE, 767--784

\bibitem[{{Riello} {et~al.}(2021){Riello}, {De Angeli}, {Evans}, {Montegriffo},
  {Carrasco}, {Busso}, {Palaversa}, {Burgess}, {Diener}, {Davidson}, {Rowell},
  {Fabricius}, {Jordi}, {Bellazzini}, {Pancino}, {Harrison}, {Cacciari}, {van
  Leeuwen}, {Hambly}, {Hodgkin}, {Osborne}, {Altavilla}, {Barstow}, {Brown},
  {Castellani}, {Cowell}, {De Luise}, {Gilmore}, {Giuffrida}, {Hidalgo},
  {Holland}, {Marinoni}, {Pagani}, {Piersimoni}, {Pulone}, {Ragaini}, {Rainer},
  {Richards}, {Sanna}, {Walton}, {Weiler}, \& {Yoldas}}]{2021A&A...649A...3R}
{Riello}, M., {De Angeli}, F., {Evans}, D.~W., {et~al.} 2021, \aap, 649, A3

\bibitem[{{Rodr{\'\i}guez Mart{\'\i}nez} {et~al.}(2023){Rodr{\'\i}guez
  Mart{\'\i}nez}, {Martin}, {Gaudi}, {Schulze}, {Asnodkar}, {Boley}, \&
  {Ballard}}]{2023AJ....166..137R}
{Rodr{\'\i}guez Mart{\'\i}nez}, R., {Martin}, D.~V., {Gaudi}, B.~S., {et~al.}
  2023, \aj, 166, 137

\bibitem[{Santerne {et~al.}(2019)Santerne, Malavolta, Kosiarek, Dai, Dressing,
  Dumusque, Hara, Lopez, Mortier, Vanderburg, {et~al.}}]{santerne2019extremely}
Santerne, A., Malavolta, L., Kosiarek, M., {et~al.} 2019, arXiv preprint
  arXiv:1911.07355

\bibitem[{Schneider {et~al.}(2011)Schneider, Dedieu, Le~Sidaner, Savalle, \&
  Zolotukhin}]{schneider2011defining}
Schneider, J., Dedieu, C., Le~Sidaner, P., Savalle, R., \& Zolotukhin, I. 2011,
  Astronomy \& Astrophysics, 532, A79

\bibitem[{Smart {et~al.}(2021)Smart, Sarro, Rybizki, Reyl{\'e}, Robin, Hambly,
  Abbas, Barstow, De~Bruijne, Bucciarelli, {et~al.}}]{smart2021gaia}
Smart, R.~L., Sarro, L., Rybizki, J., {et~al.} 2021, Astronomy \& Astrophysics,
  649, A6

\bibitem[{{Struve}(1952)}]{1952Obs....72..199S}
{Struve}, O. 1952, The Observatory, 72, 199

\bibitem[{T{\"o}rnqvist {et~al.}(1985)T{\"o}rnqvist, Vartia, \&
  Vartia}]{tornqvist1985should}
T{\"o}rnqvist, L., Vartia, P., \& Vartia, Y.~O. 1985, The American
  Statistician, 39, 43

\bibitem[{Wang(2017)}]{Wang_2017}
Wang, S. 2017, Research Notes of the {AAS}, 1, 26

\bibitem[{Weiss \& Marcy(2014)}]{weiss2014mass}
Weiss, L.~M. \& Marcy, G.~W. 2014, The Astrophysical Journal Letters, 783, L6

\bibitem[{Weiss {et~al.}(2018)Weiss, Marcy, Petigura, Fulton, Howard, Winn,
  Isaacson, Morton, Hirsch, Sinukoff, {et~al.}}]{weiss2018california}
Weiss, L.~M., Marcy, G.~W., Petigura, E.~A., {et~al.} 2018, The Astronomical
  Journal, 155, 48

\bibitem[{Weiss {et~al.}(2013)Weiss, Marcy, Rowe, Howard, Isaacson, Fortney,
  Miller, Demory, Fischer, Adams, {et~al.}}]{weiss2013mass}
Weiss, L.~M., Marcy, G.~W., Rowe, J.~F., {et~al.} 2013, The Astrophysical
  Journal, 768, 14

\bibitem[{{Weiss} {et~al.}(2022){Weiss}, {Millholland}, {Petigura}, {Adams},
  {Batygin}, {Bloch}, \& {Mordasini}}]{2022arXiv220310076W}
{Weiss}, L.~M., {Millholland}, S.~C., {Petigura}, E.~A., {et~al.} 2022, arXiv
  e-prints, arXiv:2203.10076

\bibitem[{Weiss \& Petigura(2020)}]{weiss2020kepler}
Weiss, L.~M. \& Petigura, E.~A. 2020, The Astrophysical Journal Letters, 893,
  L1

\bibitem[{Winn \& Fabrycky(2015)}]{winn2015occurrence}
Winn, J.~N. \& Fabrycky, D.~C. 2015, Annual Review of Astronomy and
  Astrophysics, 53, 409

\bibitem[{Winn {et~al.}(2018)Winn, Sanchis-Ojeda, \&
  Rappaport}]{winn2018kepler}
Winn, J.~N., Sanchis-Ojeda, R., \& Rappaport, S. 2018, New Astronomy Reviews,
  83, 37

\bibitem[{Wolfgang {et~al.}(2016)Wolfgang, Rogers, \& Ford}]{Wolfgang2016}
Wolfgang, A., Rogers, L.~A., \& Ford, E.~B. 2016, The Astrophysical Journal,
  825, 19

\bibitem[{{Yoffe} {et~al.}(2021){Yoffe}, {Ofir}, \&
  {Aharonson}}]{2021ApJ...908..114Y}
{Yoffe}, G., {Ofir}, A., \& {Aharonson}, O. 2021, \apj, 908, 114

\bibitem[{Zeng {et~al.}(2019)Zeng, Jacobsen, Sasselov, Petaev, Vanderburg,
  Lopez-Morales, Perez-Mercader, Mattsson, Li, Heising,
  {et~al.}}]{zeng2019growth}
Zeng, L., Jacobsen, S.~B., Sasselov, D.~D., {et~al.} 2019, Proceedings of the
  National Academy of Sciences, 116, 9723

\bibitem[{Zeng {et~al.}(2016)Zeng, Sasselov, \& Jacobsen}]{zeng2016mass}
Zeng, L., Sasselov, D.~D., \& Jacobsen, S.~B. 2016, The Astrophysical Journal,
  819, 127

\bibitem[{Zhu(2020)}]{zhu2020patterns}
Zhu, W. 2020, The Astronomical Journal, 159, 188

\bibitem[{{Zhu} \& {Dong}(2021)}]{2021ARA&A..59..291Z}
{Zhu}, W. \& {Dong}, S. 2021, \araa, 59, 291

\bibitem[{{Zhu} \& {Wu}(2018)}]{2018AJ....156...92Z}
{Zhu}, W. \& {Wu}, Y. 2018, \aj, 156, 92

\end{thebibliography}
\newpage

\onecolumn

\begin{appendix}
\section{Supplementary figures}
\label{sec:appendix}

\begin{figure*}[h]
\resizebox{\hsize}{!}{
    \centering
    \begin{subfigure}{}
        \includegraphics{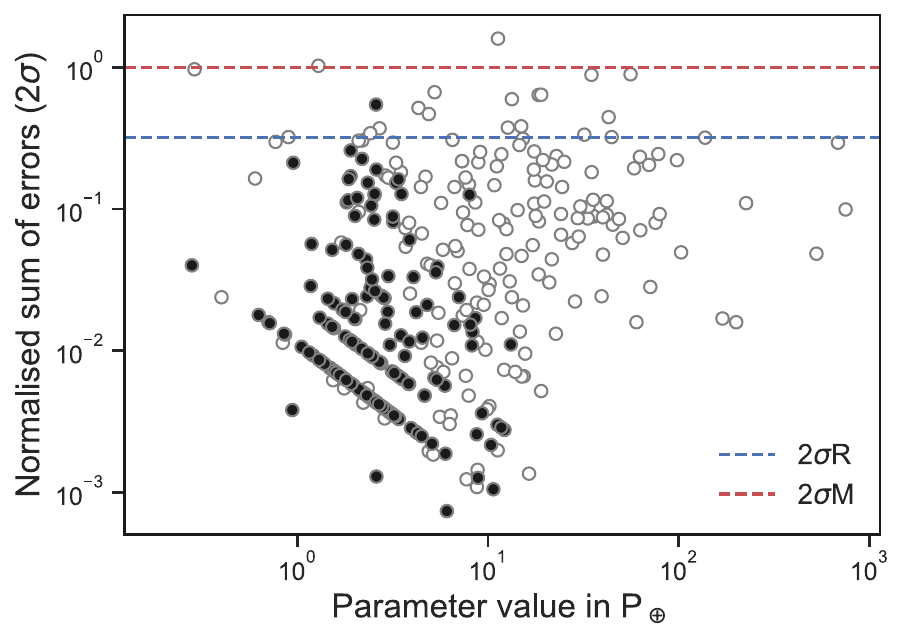}
    \end{subfigure}
    \begin{subfigure}{}
        \includegraphics{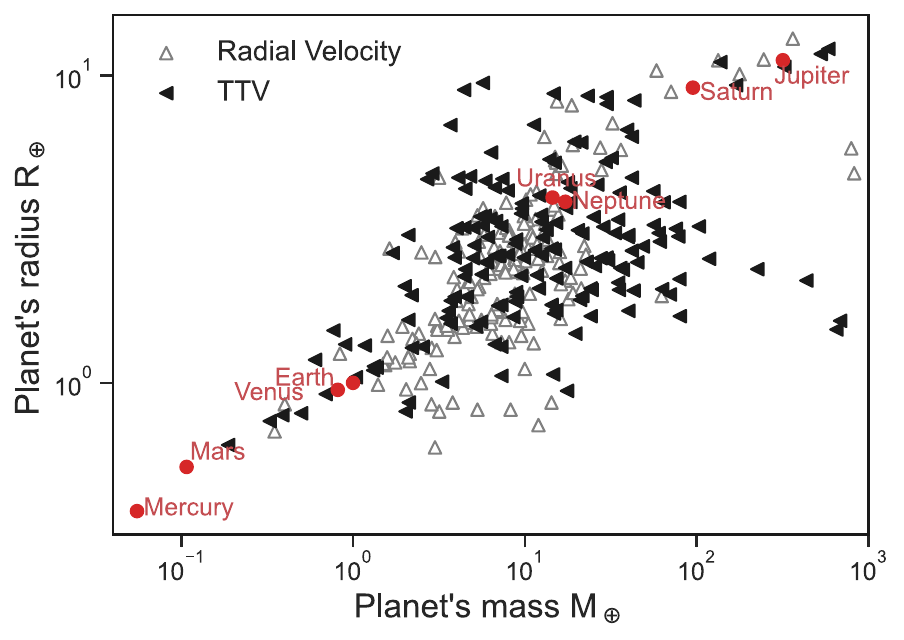}
    \end{subfigure}
    }
    \caption{Sum of errors normalised to the parameter value in the main sample (left). Black dots represent the normalised sum of of errors in radius, white dots are the same for mass. Red dashed and blue dashed lines are for the sum of errors corresponding to the construction of the sample from \citet{otegi2021similarity} for mass and radius, respectively. Planets with mass determined from RV (white upwards-pointing triangles) and TTVs (black left-pointing triangles) methods in the main sample.}
    \label{fig:33}
\end{figure*}

\begin{figure}[h]\resizebox{\hsize}{!}{
    \centering
    \begin{subfigure}
        \centering
        \includegraphics{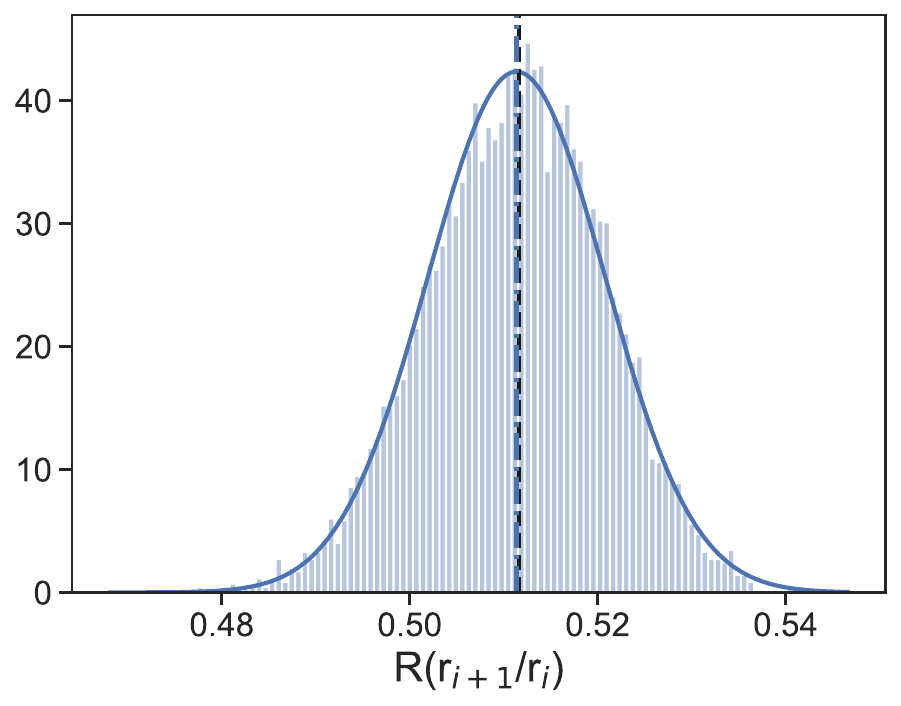}
    \end{subfigure}
    \begin{subfigure}
        \centering
        \includegraphics{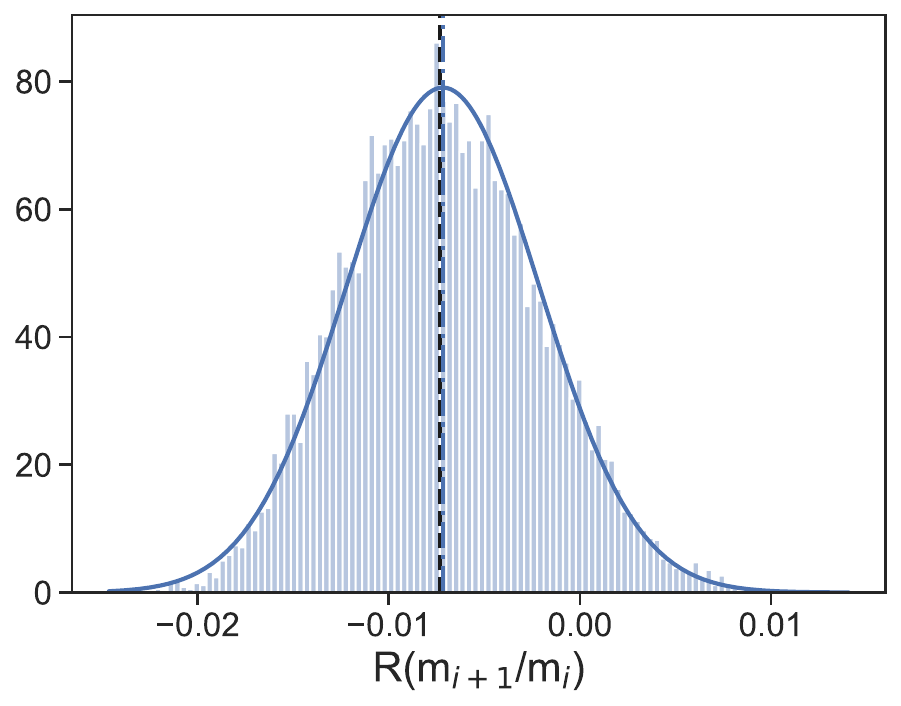}
    \end{subfigure}
    \begin{subfigure}
        \centering
        \includegraphics{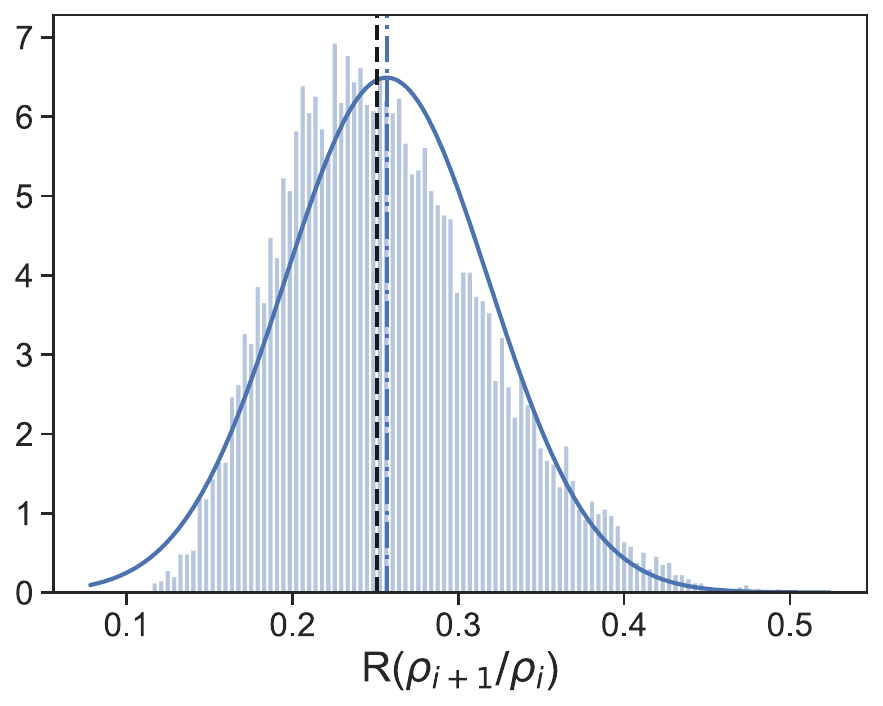}
    \end{subfigure}
    \begin{subfigure}
        \centering
        \includegraphics{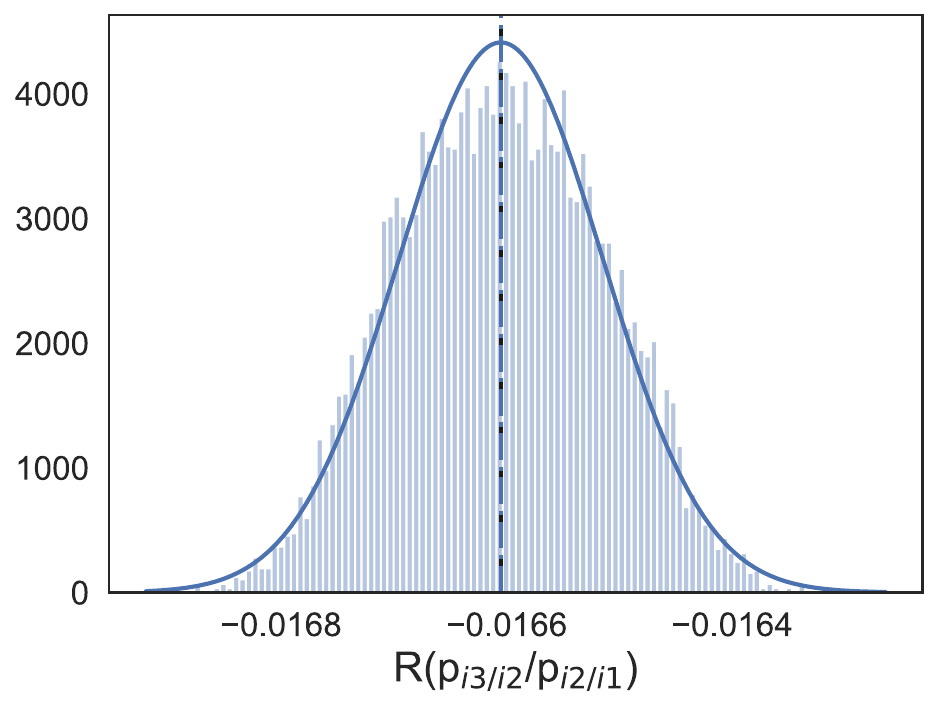}
    \end{subfigure}
    }
    \caption{Pearson coefficients distributions from the error-accommodation method. Each graph shows R distribution for parameters of adjacent planets, which were tested for linear correlation by running 10$^5$ random uniform simulations. From left to right the correlation is calculated for radii, masses, densities of adjacent planets in pairs and period ratios in triples. The blue dash-dotted and black dashed lines indicate mean and median values, respectively. The solid blue line indicates the best-fitted probability density function (PDF) of the normal distribution. The y-axis shows the probability density (the probability per unit on the x-axis).}
    \label{fig:23a}
\end{figure}
\begin{figure}[h]\resizebox{\hsize}{!}{
    \centering
    \begin{subfigure}
        \centering
        \includegraphics{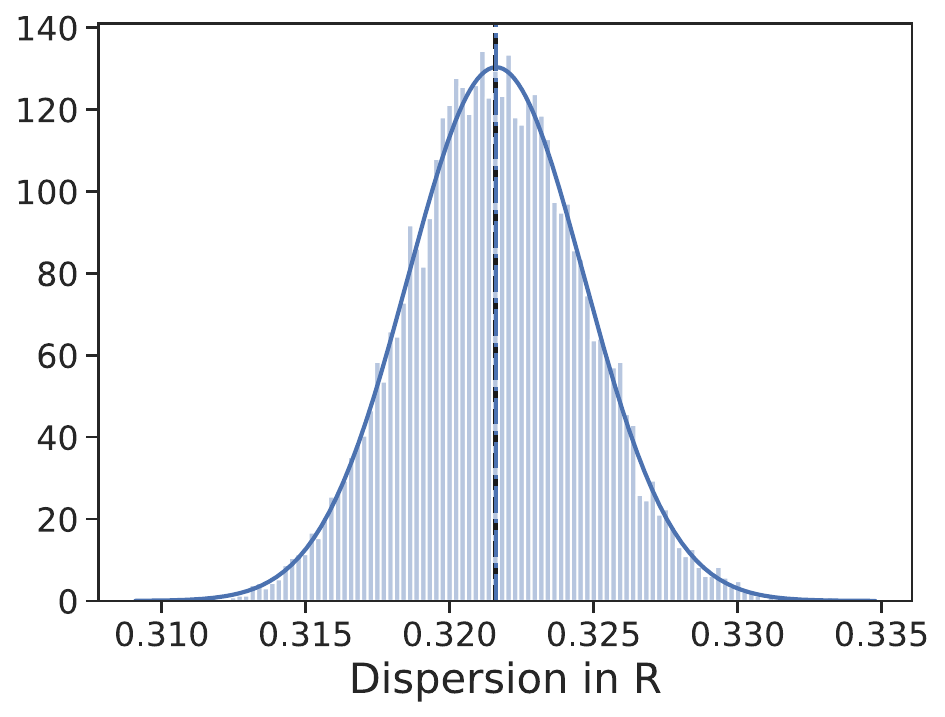}
    \end{subfigure}
    \begin{subfigure}
        \centering
        \includegraphics{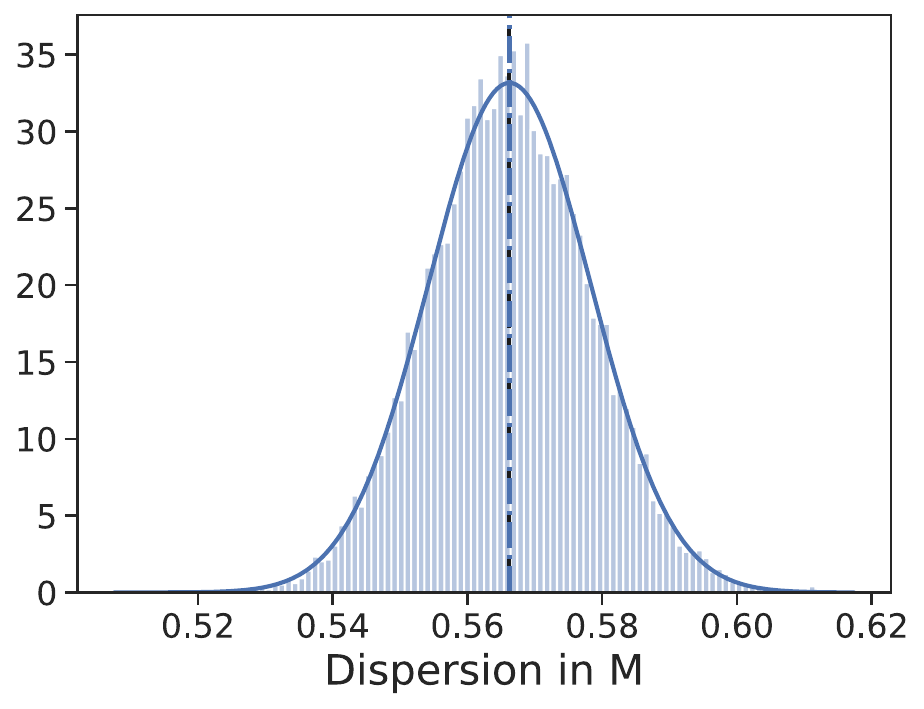}
    \end{subfigure}
    \begin{subfigure}
        \centering
        \includegraphics{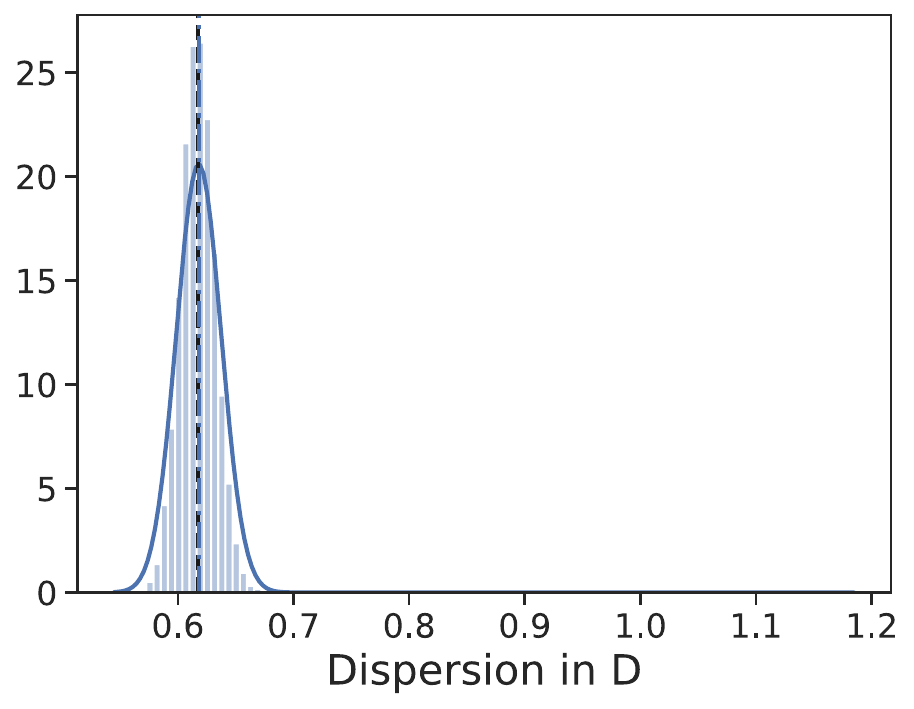}
    \end{subfigure}
    \begin{subfigure}
        \centering
        \includegraphics{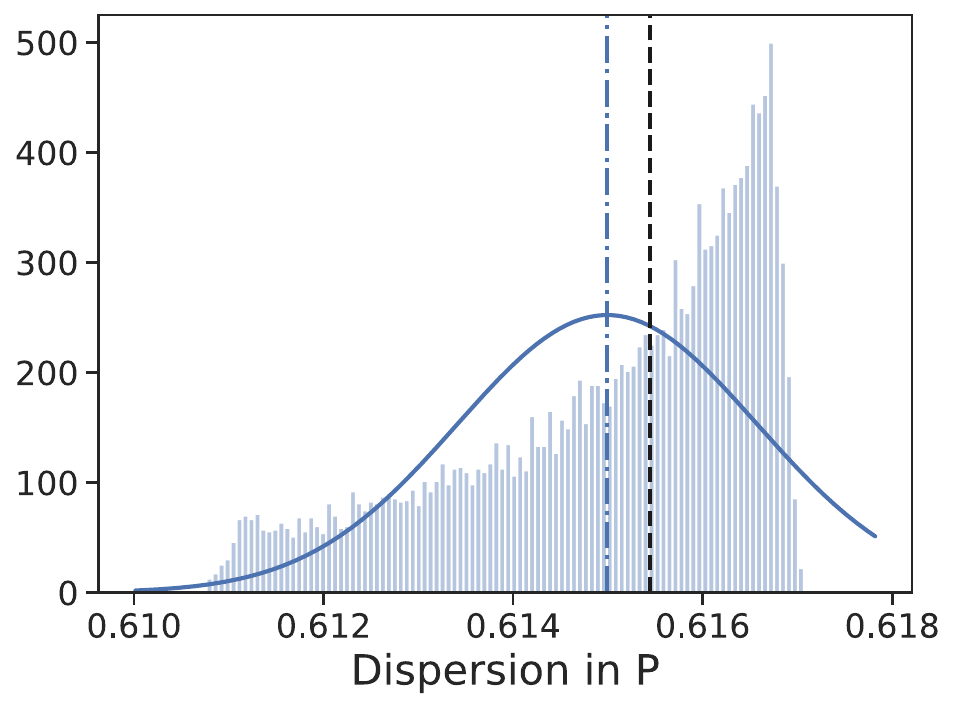}
    \end{subfigure}
    }
    \caption{Dispersion in parameters from the error-accommodation method. Each graph shows intrasystem dispersion distribution in the main sample acquired by running 10$^5$ random uniform simulations within the error intervals. From left to right the distributions are obtained for radii, masses, densities of adjacent planets in systems with at least two planets and period ratios in systems with at least three planets. The blue dash-dotted and black dashed lines indicate mean and median values, respectively. The solid blue line indicates the best-fitted probability density function (PDF) of the normal distribution. The y-axis shows the probability density (the probability per unit on the x-axis).}
    \label{fig:28a}
\end{figure}
\begin{figure}[h]\resizebox{\hsize}{!}{
    \centering
    \begin{subfigure}
        \centering
        \includegraphics{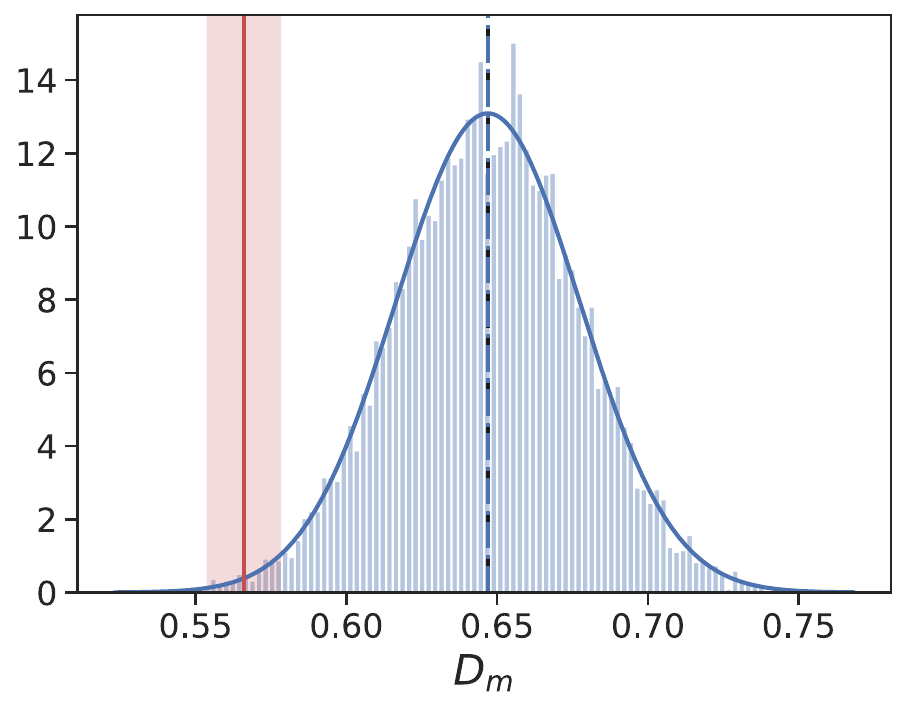}
    \end{subfigure}
    \begin{subfigure}
        \centering
        \includegraphics{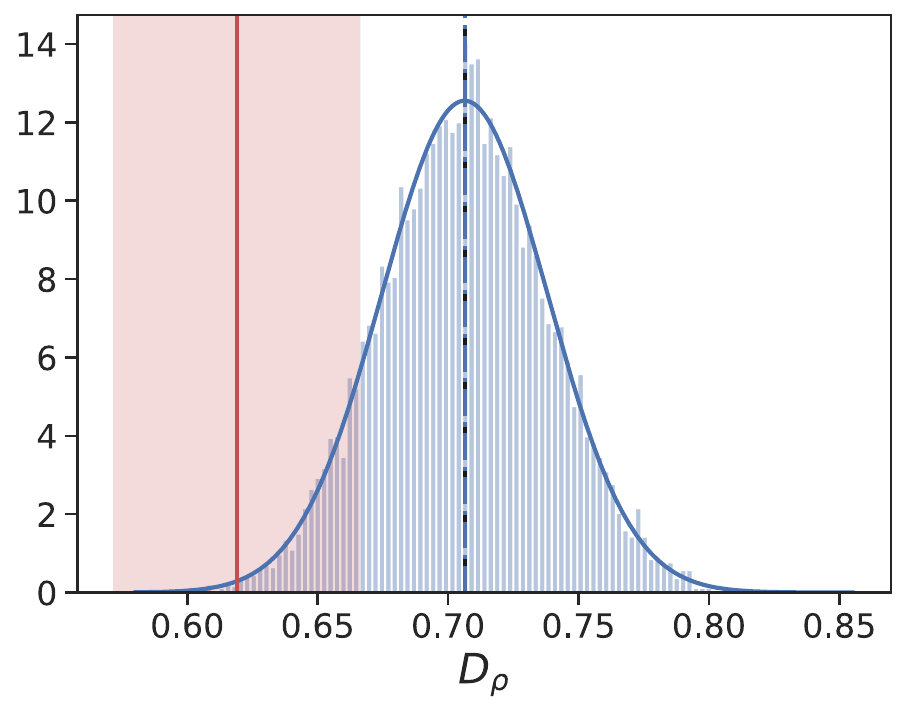}
    \end{subfigure}
    \begin{subfigure}
        \centering
        \includegraphics{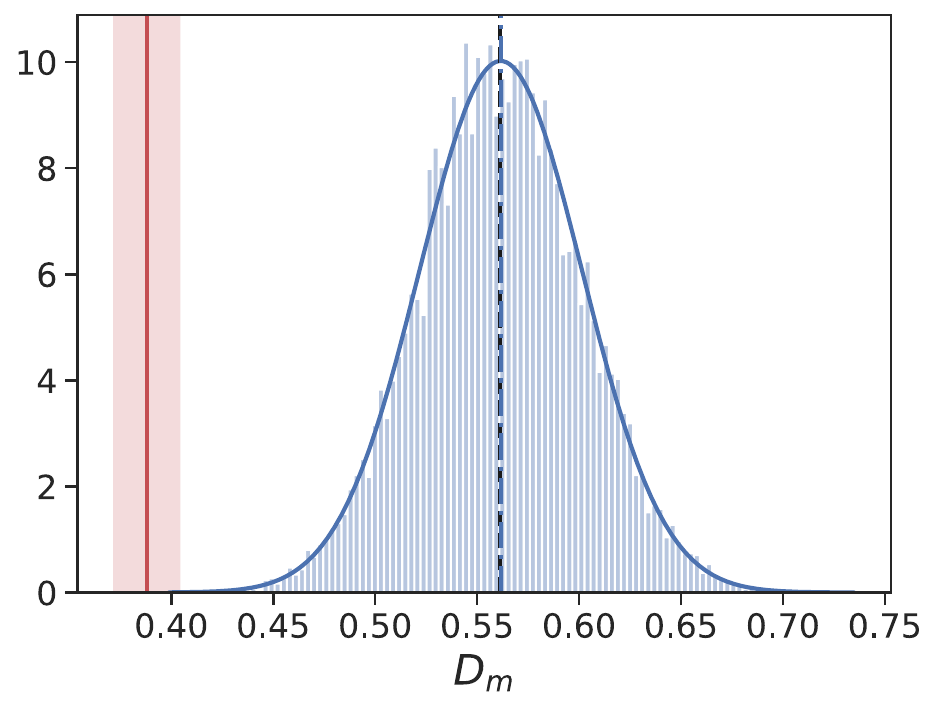}
    \end{subfigure}
    \begin{subfigure}
        \centering
        \includegraphics{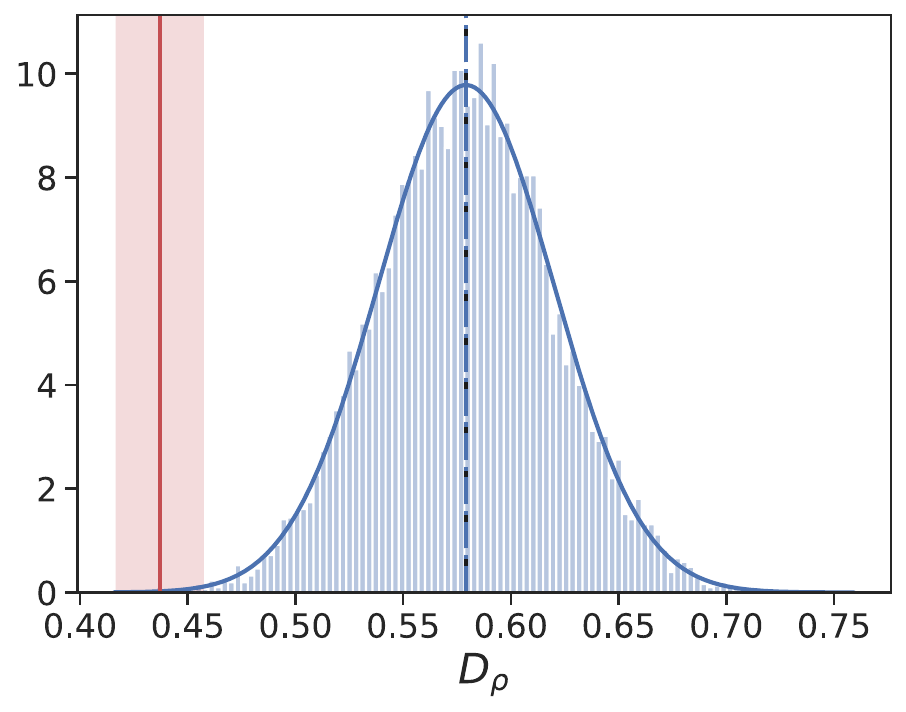}
    \end{subfigure}
    }
    \caption{Expected dispersion from the non-parametric MRR \citep{2018ApJ...869....5N} compared to dispersion in mass and density from the error-accommodation method. Red solid lines show intrasystem dispersion $D_{err}$ and the corresponding standard deviation (light-red shadowed regions). From left to right the 10$^5$ random distributions from MRR are obtained for masses, and densities of adjacent planets in the main sample, and for masses, and densities of adjacent planets in the HCR sample. The blue dash-dotted and black dashed lines indicate mean and median values, respectively. The solid blue line indicates the best-fitted probability density function (PDF) of the normal distribution. The y-axis shows the probability density (the probability per unit on the x-axis).}
    \label{fig:28c}
\end{figure}

\begin{figure}[h]
    \centering
    \begin{subfigure}{}
        \includegraphics[width=1\linewidth]{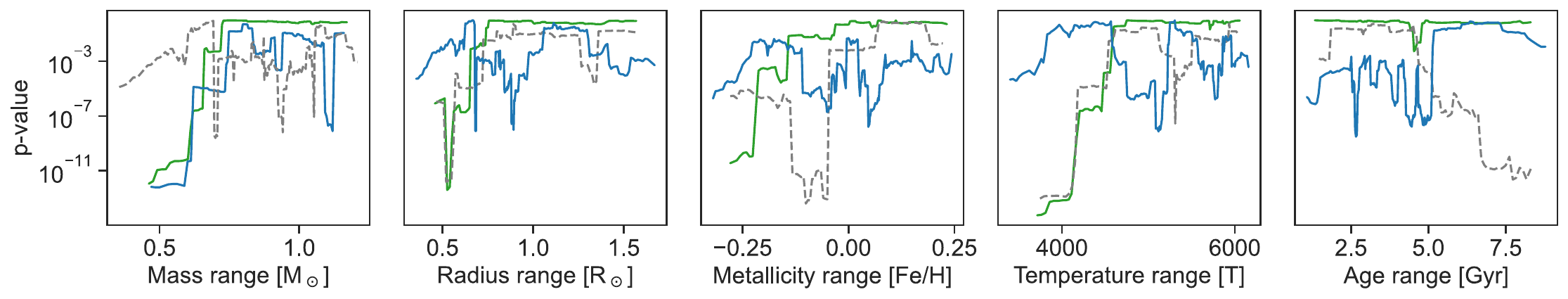}
    \end{subfigure}
    \begin{subfigure}{}
        \includegraphics[width=1\linewidth]{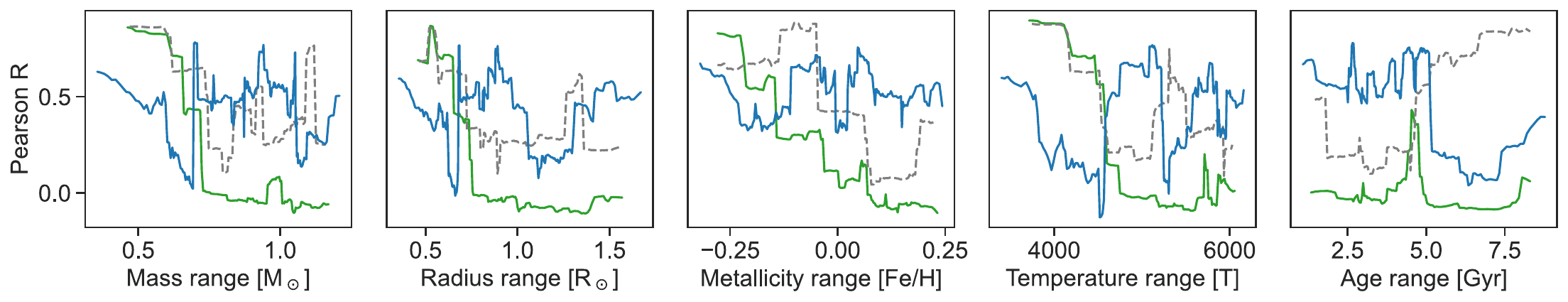}
    \end{subfigure}
    \caption{Moving window test of the "40 data points" moving subsamples for masses (green line), radii (blue line), and densities (grey dashed line) of adjacent planets pairs. R and p-values from the repeated Pearson correlation analyses are plotted against ranges of stellar parameters. The values of the certain stellar parameter on the x-axis are calculated as the mean values for all planets in each subsample.}
    \label{fig:23}
\end{figure}

\begin{figure}[h]\resizebox{\hsize}{!}{
    \centering
    \begin{subfigure}
        \centering
        \includegraphics{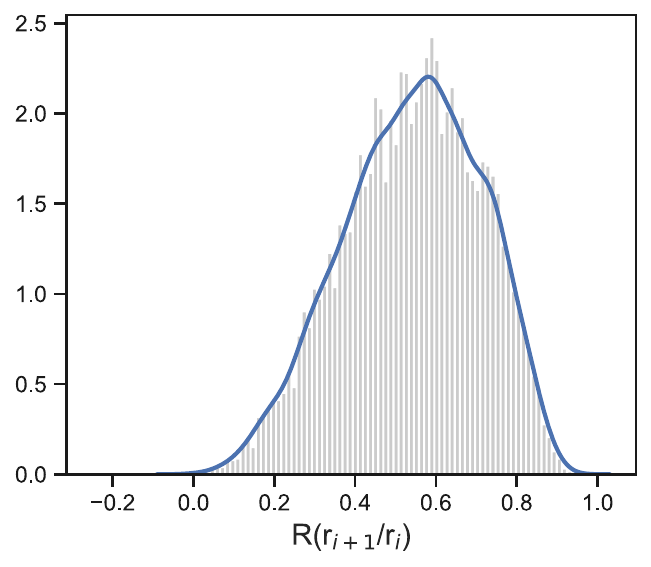}
    \end{subfigure}
    \begin{subfigure}
        \centering
        \includegraphics{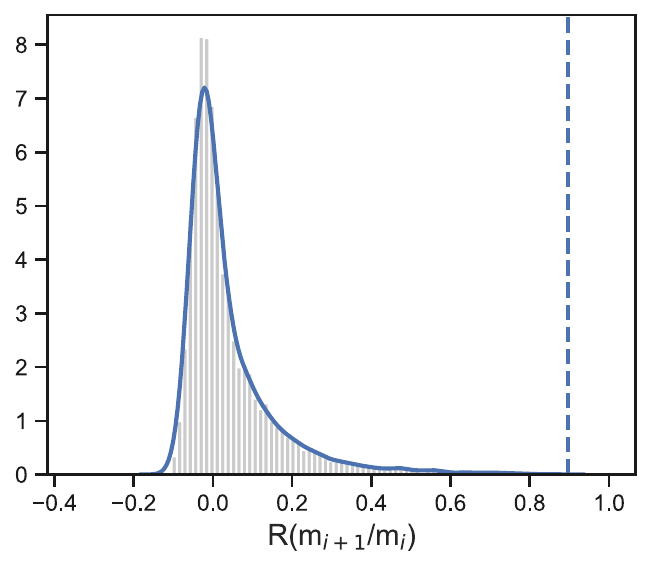}
    \end{subfigure}
    \begin{subfigure}
        \centering
        \includegraphics{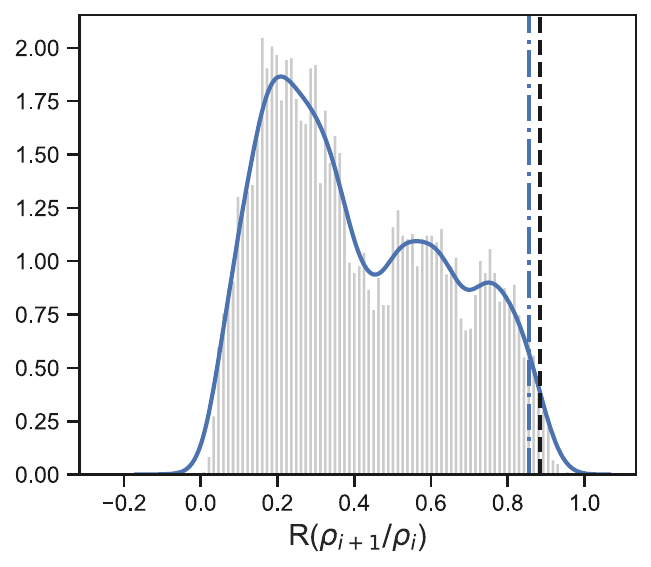}
    \end{subfigure}
    \begin{subfigure}
        \centering
        \includegraphics{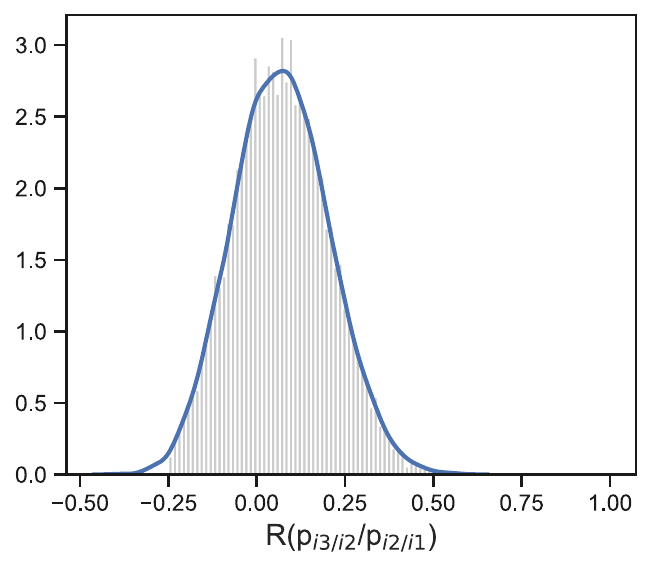}
    \end{subfigure}
    }
    \caption{Bootstrap test statistics. Each graph shows the Pearson coefficient distribution for parameter pairs for bootstrapping 40 planet pairs in 10$^5$ simulations. From left to right the correlation is calculated for radii, masses, densities of adjacent planets in pairs and period ratios in triples. In the "mass" graph the blue dashed line indicates the "high-correlated" subsample for planets around cool stars (2566-4398 K). In the "density" graph the blue dash-dotted and black dashed lines indicate the "high-correlated" subsamples for planets around stars with metallicity ([Fe/H] $\in$[-0.210; -0.010]) and old stars (6.03-12.3 Gyr) respectively.}
    \label{fig:29}
\end{figure}

\begin{figure*}[ht!]\resizebox{\hsize}{!}{

   \includegraphics[width=18cm]{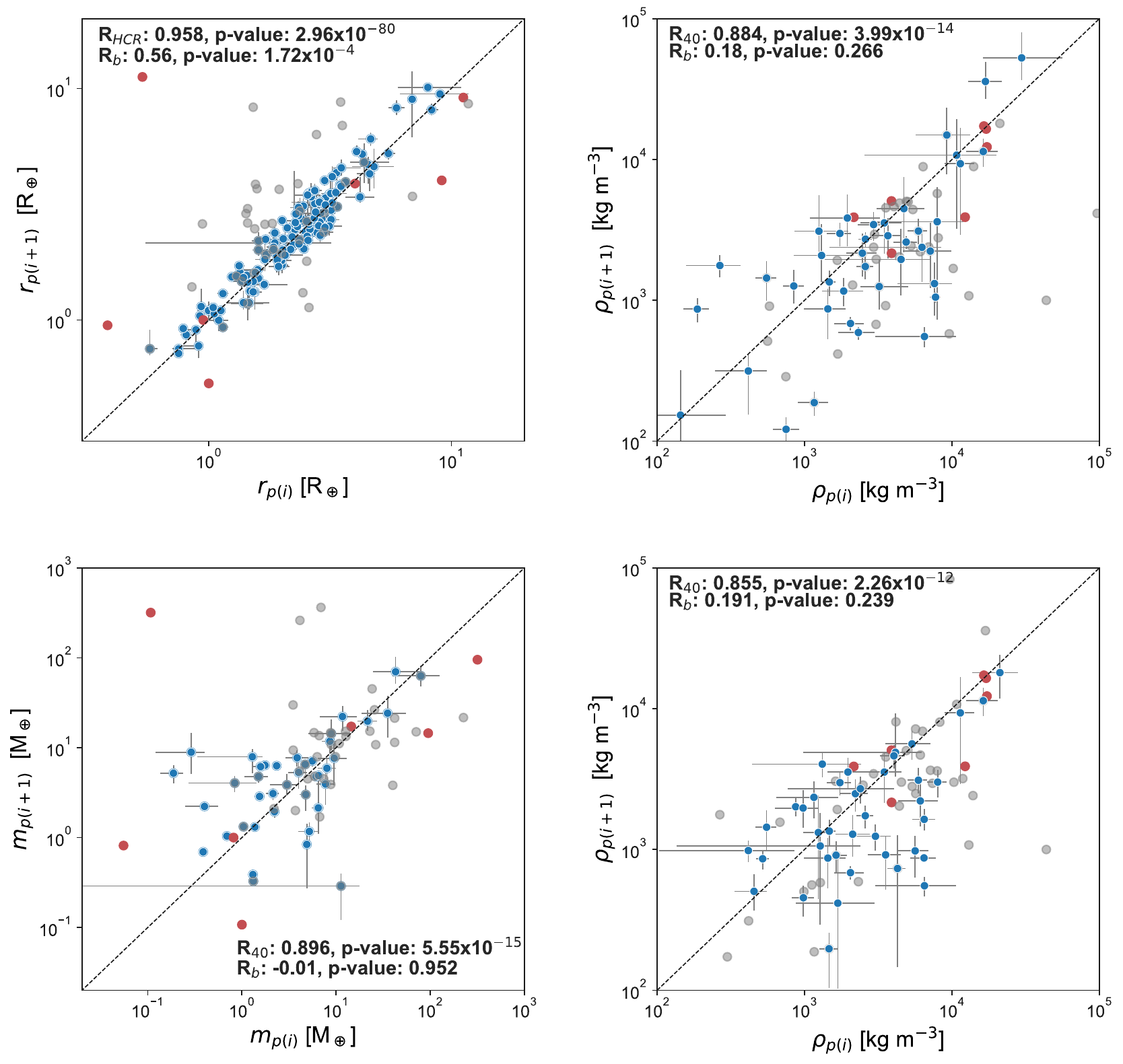}
}
    \caption{HCR (149 pairs), and stellar 40-pair HCM and HCD subsamples. The HCR sample and comparison to a random bootstrap of 40 pairs (top left).The planet pairs (blue dots) around host stars with similar properties: around cool stars, T$_\mathrm{eff} \in$[2566; 4398] K (masses, bottom left), around those with [Fe/H] $\in$[-0.210; -0.010] and old stars, age $\in$[6.03; 12.3] Gyr (top and bottom right, densities).  The grey dots represent one realisation of the bootstrapped 40-pair samples out of 10$^5$ simulations. The red dots correspond to the Solar System planets. The black dashed line is the 1:1 line.}
\label{fig:28b}
\end{figure*}
\end{appendix}
\end{document}